\documentclass[%
 reprint,
preprintnumbers,
 amsmath,amssymb,
 aps,
prd,
 superscriptaddress,
]{revtex4-2}

\usepackage{graphicx}
\usepackage{dcolumn}
\usepackage{multirow}
\usepackage{bm}
\usepackage{hyperref}
\begin{document}

\title{Scintillation yield from electronic and nuclear recoils in superfluid $^4$He}

\author{A. Biekert} \affiliation{University of California Berkeley, Department of Physics, Berkeley, CA 94720, USA}
\author{C.~Chang} \affiliation{Argonne National Laboratory, 9700 S Cass Ave, Lemont, IL 60439, USA}
\author{C.~W.~Fink} \affiliation{University of California Berkeley, Department of Physics, Berkeley, CA 94720, USA} 
\author{M.~Garcia-Sciveres} \affiliation{Lawrence Berkeley National Laboratory, 1 Cyclotron Rd., Berkeley, CA 94720, USA} 
\author{E.~C.~Glazer} \affiliation{University of California Berkeley, Department of Physics, Berkeley, CA 94720, USA}
\author{W.~Guo} \affiliation{Department of Mechanical Engineering, FAMU-FSU College of Engineering, Florida State University, Tallahassee, FL 32310, USA} \affiliation{National High Magnetic Field Laboratory, Tallahassee, FL 32310, USA}
\author{S.A.~Hertel} \affiliation{University of Massachusetts, Amherst Center for Fundamental Interactions and Department of Physics, Amherst, MA 01003-9337 USA}
\author{S.~Kravitz} \affiliation{Lawrence Berkeley National Laboratory, 1 Cyclotron Rd., Berkeley, CA 94720, USA} 
\author{J.~Lin} \thanks{Corresponding author: \href{mailto:junsonglin@berkeley.edu}{junsonglin@berkeley.edu}} \affiliation{University of California Berkeley, Department of Physics, Berkeley, CA 94720, USA}
\author{M.~Lisovenko} \affiliation{Argonne National laboratory, 9700 S Cass Ave, Lemont, IL 60439, USA}
\author{R.~Mahapatra} \affiliation{Texas A\&M University, Department of Physics and Astronomy, College Station, TX 77843-4242, USA}
\author{D.~N.~McKinsey} \affiliation{University of California Berkeley, Department of Physics, Berkeley, CA 94720, USA} \affiliation{Lawrence Berkeley National Laboratory, 1 Cyclotron Rd., Berkeley, CA 94720, USA} 
\author{J.~S.~Nguyen} \affiliation{University of California Berkeley, Department of Physics, Berkeley, CA 94720, USA}
\author{V.~Novosad} \affiliation{Argonne National laboratory, 9700 S Cass Ave, Lemont, IL 60439, USA}
\author{W.~Page} \affiliation{University of California Berkeley, Department of Physics, Berkeley, CA 94720, USA}
\author{P.~K.~Patel} \affiliation{University of Massachusetts, Amherst Center for Fundamental Interactions and Department of Physics, Amherst, MA 01003-9337 USA}
\author{B.~Penning} \affiliation{University of Michigan, Randall Laboratory of Physics, Ann Arbor, MI 48109-1040, USA} 
\author{H.~D.~Pinckney} \affiliation{University of Massachusetts, Amherst Center for Fundamental Interactions and Department of Physics, Amherst, MA 01003-9337 USA}
\author{M.~Pyle} \affiliation{University of California Berkeley, Department of Physics, Berkeley, CA 94720, USA}
\author{R.~K.~Romani} \affiliation{University of California Berkeley, Department of Physics, Berkeley, CA 94720, USA}
\author{A.~S.~Seilnacht} \affiliation{University of California Berkeley, Department of Physics, Berkeley, CA 94720, USA}
\author{A.~Serafin} \affiliation{University of Massachusetts, Amherst Center for Fundamental Interactions and Department of Physics, Amherst, MA 01003-9337 USA}
\author{R.~J.~Smith} \affiliation{University of California Berkeley, Department of Physics, Berkeley, CA 94720, USA} 
\author{P.~Sorensen} \affiliation{Lawrence Berkeley National Laboratory, 1 Cyclotron Rd., Berkeley, CA 94720, USA} 
\author{B.~Suerfu} \affiliation{University of California Berkeley, Department of Physics, Berkeley, CA 94720, USA} 
 \author{A.~Suzuki} \affiliation{Lawrence Berkeley National Laboratory, 1 Cyclotron Rd., Berkeley, CA 94720, USA}
\author{V.~Velan} \affiliation{University of California Berkeley, Department of Physics, Berkeley, CA 94720, USA}  
\author{G.~Wang} \affiliation{Argonne National laboratory, 9700 S Cass Ave, Lemont, IL 60439, USA}
\author{S.~L.~Watkins} \affiliation{University of California Berkeley, Department of Physics, Berkeley, CA 94720, USA}
\author{V.~G.~Yefremenko} \affiliation{Argonne National laboratory, 9700 S Cass Ave, Lemont, IL 60439, USA}
\author{L.~Yuan} \affiliation{University of California Berkeley, Department of Physics, Berkeley, CA 94720, USA}
\author{J.~Zhang} \affiliation{Argonne National laboratory, 9700 S Cass Ave, Lemont, IL 60439, USA}

\collaboration{SPICE/HeRALD Collaboration}

\date{\today}

\begin{abstract}
Superfluid $^4$He is a promising target material for direct detection of light ($<$ 1 GeV) dark matter. Possible signal channels available for readout in this medium include prompt photons, triplet excimers, and roton and phonon quasiparticles.  The relative yield of these signals has implications for the sensitivity and discrimination power of a superfluid $^4$He dark matter detector. Using a 16~cm$^3$ volume of 1.75~K superfluid $^4$He read out by six immersed photomultiplier tubes, we measured the scintillation from electronic recoils ranging between 36.3 and 185~keV$_\mathrm{ee}$, yielding a mean signal size of $1.25^{+0.03}_{-0.03}$~phe/keV$_\mathrm{ee}$, and nuclear recoils from 53.2 to 1090~keV$_\mathrm{nr}$. We compare the results of our relative scintillation yield measurements to an existing semiempirical model based on helium-helium and electron-helium interaction cross sections. We also study the behavior of delayed scintillation components as a function of recoil type and energy, a further avenue for signal discrimination in superfluid $^4$He.
\end{abstract}

\maketitle

\section{Introduction}
Superfluid $^4$He has been proposed for use in dark matter direct detection \cite{Adams:1996,Guo:13prd,Ito13,Mar:2017,Hertel:2019,JL2021a}. In particular, low-mass dark matter models have received substantial recent development and interest \cite{Alexander:2016aln,Bat:2017, brn_report}. Superfluid helium has potential to allow particle identification through measurement of the ratio of charge to light \cite{Guo:13prd}, the ratio of electronic to quasiparticle excitations \cite{Hertel:2019}, or the scintillation pulse shape due to the timing of different components \cite{McKinsey:03pra}, enabling lowered backgrounds and increased discovery potential. Liquid helium scintillation has been used in ultracold neutron experiments \cite{Huffman:00n} and was proposed for measurement of the $pp$ solar neutrino flux \cite{Lanou:87prl, McKinsey:2000}. Understanding the scintillation yields and time dependencies for both nuclear and electronic recoils will aid in the design and implementation of future dark matter direct detection experiments, as well as future searches for the permanent electric dipole moment of the neutron \cite{Golub:94pr,Ahmed:2019} and measurements of coherent elastic neutrino-nucleus scattering \cite{Scholberg:06prd,Cadeddu:2019}.

Energy deposition in superfluid helium results in a rich variety of excited species, and the relative population of these species contains information about the primary track. Signals in superfluid helium include singlet diatomic molecules (excimers), triplet excimers, rotons, phonons, and perhaps quantum turbulence \cite{adamsthesis, Dodd:98, Forrester:2013}. Recent experimental measurements of superfluid helium response signals have focused on scintillation as a function of temperature and applied electric field from $\alpha$ and electron sources \cite{TI2012a, Guo:12ji, GS2014a, NP2020a}. In addition to directly measuring these signal channels, predictions of helium response to nuclear and electronic recoils can be constructed from experimental data of He-He and electron-He interaction cross sections \cite{Guo:13prd, Ito13}, yielding a full model of the partitioning into various signal channels as a function of recoil energy \cite{GS2016a, Hertel:2019, GS2019a}.

The signal channels in superfluid helium can also be probed by analyzing the temporal distribution of scintillation. Early experiments on superfluid helium scintillation revealed significant emission of delayed photons over one microsecond after the initial excitation \cite{fischbach69}. Scintillation signals in the first millisecond following a recoil include a prompt component ($<$ 10 ns), an exponentially decaying component with a lifetime of 1.6~$\mu$s, and a component that decays roughly as $t^{-1}$ \cite{McKinsey:03pra}. The $t^{-1}$ component is generally attributed to Penning ionization of triplet states, which can recombine into singlets. The source of the exponential component is less certain; in Ref.~\cite{McKinsey:03pra} it was hypothesized to be due to a reaction between a metastable state such as He($2^{1}S$) and the ground state helium atoms. It was also observed that the $t^{-1}$ component was a larger fraction of the total scintillation for an $\alpha$ source and for $^3$He neutron capture events than for a $\beta$ source. 

In this article, we report measurements of superfluid helium scintillation and its time dependence for both electronic and nuclear recoils. Section~\ref{sec:hardware} describes the experimental apparatus, Sec.~\ref{sec:simulations} describes Monte Carlo simulations used in the analysis, Sec.~\ref{sec:analysis} describes the data analysis, Sec.~\ref{sec:discussion} discusses and summarizes the results, and Sec.~\ref{sec:conclusion} contains concluding remarks. 

\section{Experimental apparatus}
\label{sec:hardware}
\subsection{Helium detector hardware}

\begin{figure}[ht]
\includegraphics[width=0.75\linewidth]{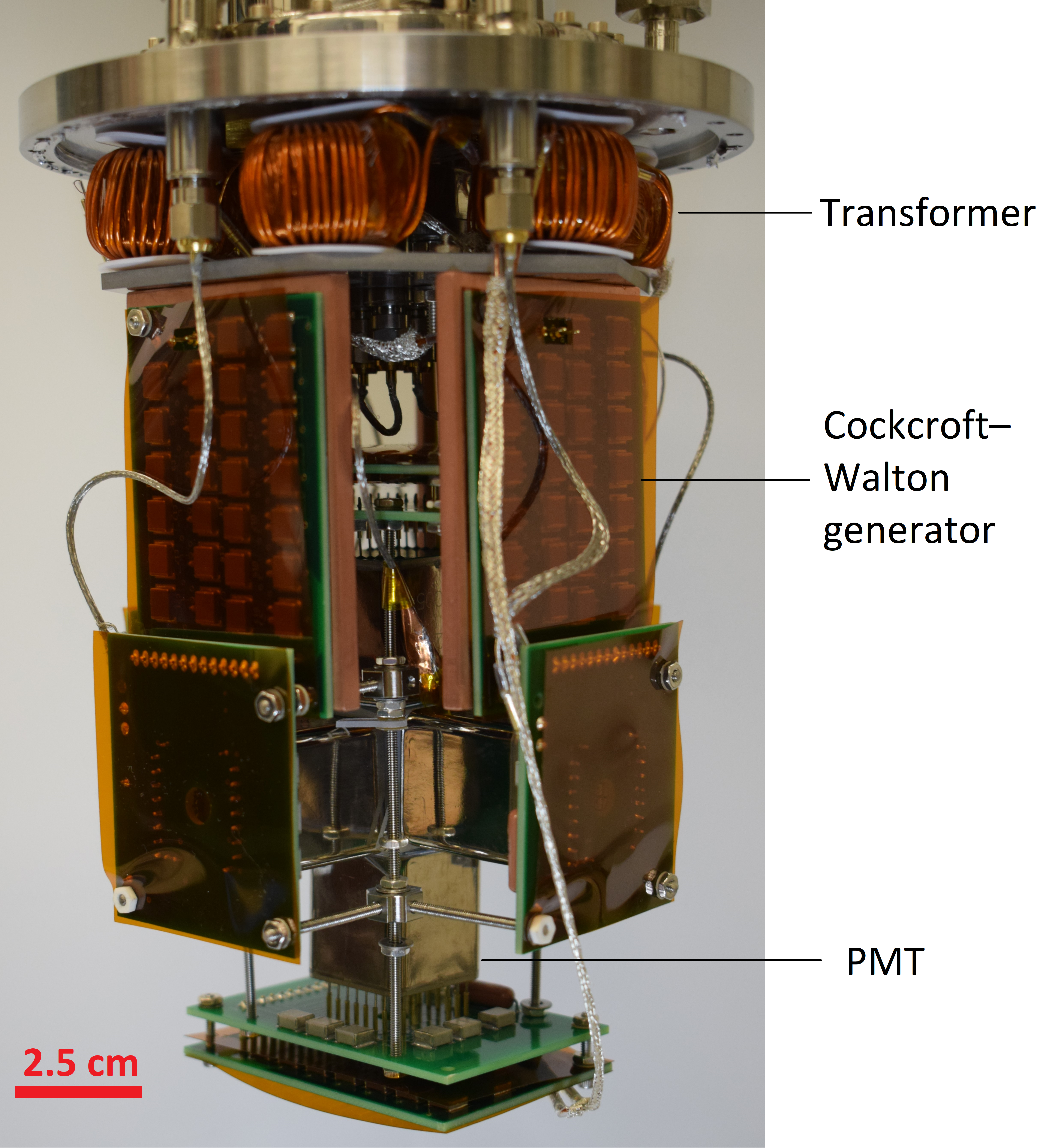}
\caption{\label{fig:Detector_photo} Photograph of the detector used in this work.}
\end{figure}

\begin{figure}[ht]
\includegraphics[width=0.5\linewidth]{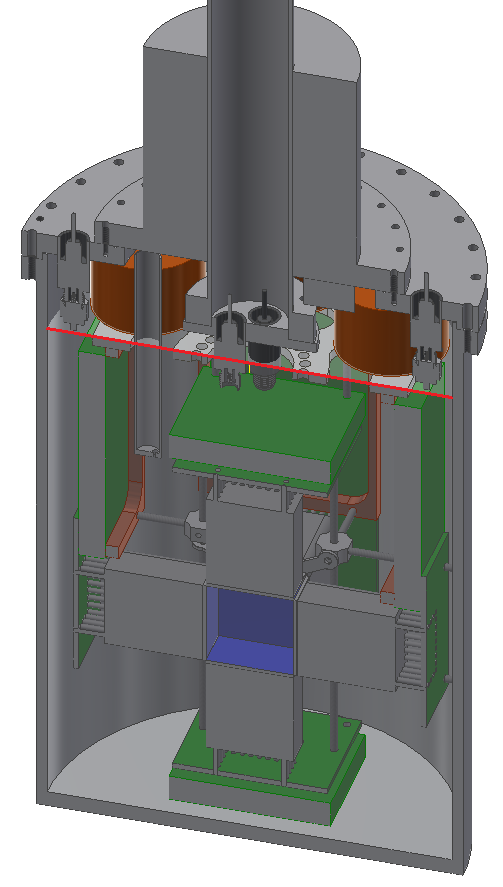}
\caption{\label{fig:Detector_CAD} CAD model of the detector. The TPB wavelength shifter is shown in blue, defining the active volume of the detector. The measured $^4$He liquid-gas interface at 1.75 K is indicated by a red line, near the top of the green printed circuit boards.}
\end{figure}

The detector consisted of six Hamamatsu R8520-06-MOD photomultiplier tubes (PMTs), with cathode platinum underlay, monitoring a 16~cm$^3$ cubic volume of superfluid $^4$He, as shown in Figs.~\ref{fig:Detector_photo} and \ref{fig:Detector_CAD}. Because the PMT windows are not transparent to the 80~nm wavelength prompt scintillation light of $^4$He, tetraphenyl butadiene (TPB) \cite{MCKINSEY1997351, Benson2018} was used to first shift the light to 430~nm.  The TPB was deposited to a thickness of 0.3 mg/cm\textsuperscript{2} on 1~mm thick fused quartz plates, arranged in a cubic geometry illustrated in Fig.~\ref{fig:Detector_CAD}.

The cryogenic PMT readout, previously developed for use at milli-Kelvin temperatures in vacuum \cite{ZHANG2016}, was adapted for the superfluid $^4$He environment in this work. Each PMT was biased by an individual Cockcroft–Walton generator (CWG). The PMT dynode stages were connected to the different voltage levels produced by the CWG without resistive voltage divider circuits. In this way, the heat load was minimized and no cryogenic high-voltage feedthroughs were needed. External 3~V peak-to-peak AC with 26 kHz frequency was first amplified by transformers with a winding ratio 1:24 and a molypermalloy powder core, then fed to the CWG. PMT gains were monitored with the slow scintillation component from superfluid $^4$He, which we describe in more detail in Sec.~\ref{sssec:SPE_size}.

The whole detector was mounted inside a pumped $^4$He cryostat, manufactured by Janis Cryogenics, with a 1~K pot. The sample-space $^4$He, nearly filling the chamber shown in Fig.~\ref{fig:Detector_CAD}, was separated from the cooling $^4$He.

\subsection{Cool down process}
While helium permeation into PMTs is generally a concern, the permeability drops exponentially with temperature \cite{Hyodo1980}. Thus, a special cooldown procedure was used to prevent helium permeation into the PMTs in this experiment. The cryostat was first cooled to liquid nitrogen temperature (77 K); the sample space was filled with high purity nitrogen heat exchange gas during this step. Next, the sample space was flushed with high purity helium gas (impurities $<$200 ppb) six times to ensure all of the nitrogen exchange gas was removed. We estimate less than 10\textsuperscript{-10} bar nitrogen was left after this flushing procedure. The sample space was then filled with 1 bar of high purity helium exchange gas, and the cryostat was further cooled down to 4.2 K. High purity $^4$He was condensed into the sample space by the cooling power of the 1~K pot, after it was further purified by a 4.2 K cold trap. Once the sample space was completely filled with liquid $^4$He at 4.2 K, it was cooled down through the superfluid phase transition to a final temperature of 1.75 K. The data  taking  took  50  days  with  the  same  fill  of  sample helium in the detector and the entry valve at room temperature to the sample space shut. The temperature was stabilized to $~\pm$0.1 K within 1.75 K during the data taking.

\subsection{Radioactive sources and data taking configuration}
We follow existing conventions by defining the recoil energy from electronic recoils (ERs), $E_{ee}$, in units of keV$_\mathrm{ee}$, where ``ee" stands for electron equivalent. Similarly, we define the nuclear recoil (NR) energy, $E_{nr}$, in units of keV$_\mathrm{nr}$, where ``nr" stands for nuclear recoil. For a given scintillation signal strength, $E_{nr}$ is generally larger than $E_{ee}$ because NRs generate electronic excitations less efficiently than ERs do. We expect $E_{nr}$ to scale with signal strength in a nonlinear fashion, in contrast to the expected linear scaling of $E_{ee}$. 

ER data were obtained using a 270~kBq $^{137}$Cs source, which emits a gamma ray of 661.7~keV. Two cylindrical 5.1~cm diameter, 5.1~cm thick NaI detectors with PMT readout were used as far side detectors to tag gamma rays which Compton scattered in the helium target at specific angles. NR data were obtained using the 2.8~MeV neutrons from a Thermo Scientific MP 320 deuterium-deuterium (DD) fusion neutron generator producing roughly $1 \times 10^6$~neutrons/second in all directions. A 12.7~cm diameter, 12.7~cm thick cylindrical BC-501A liquid scintillator (LS) detector with PMT readout was used to tag neutrons recoiling in the helium detector at a specific angle. Laser tools were used to measure the positions and orientation of the radioactive sources and the detectors, with an estimated position uncertainty of 1~mm. 

\subsection{Data acquisition and trigger efficiency}
\label{ssec:data_acq_and_trigger_eff}
\begin{figure}[ht]
\includegraphics[width=\linewidth]{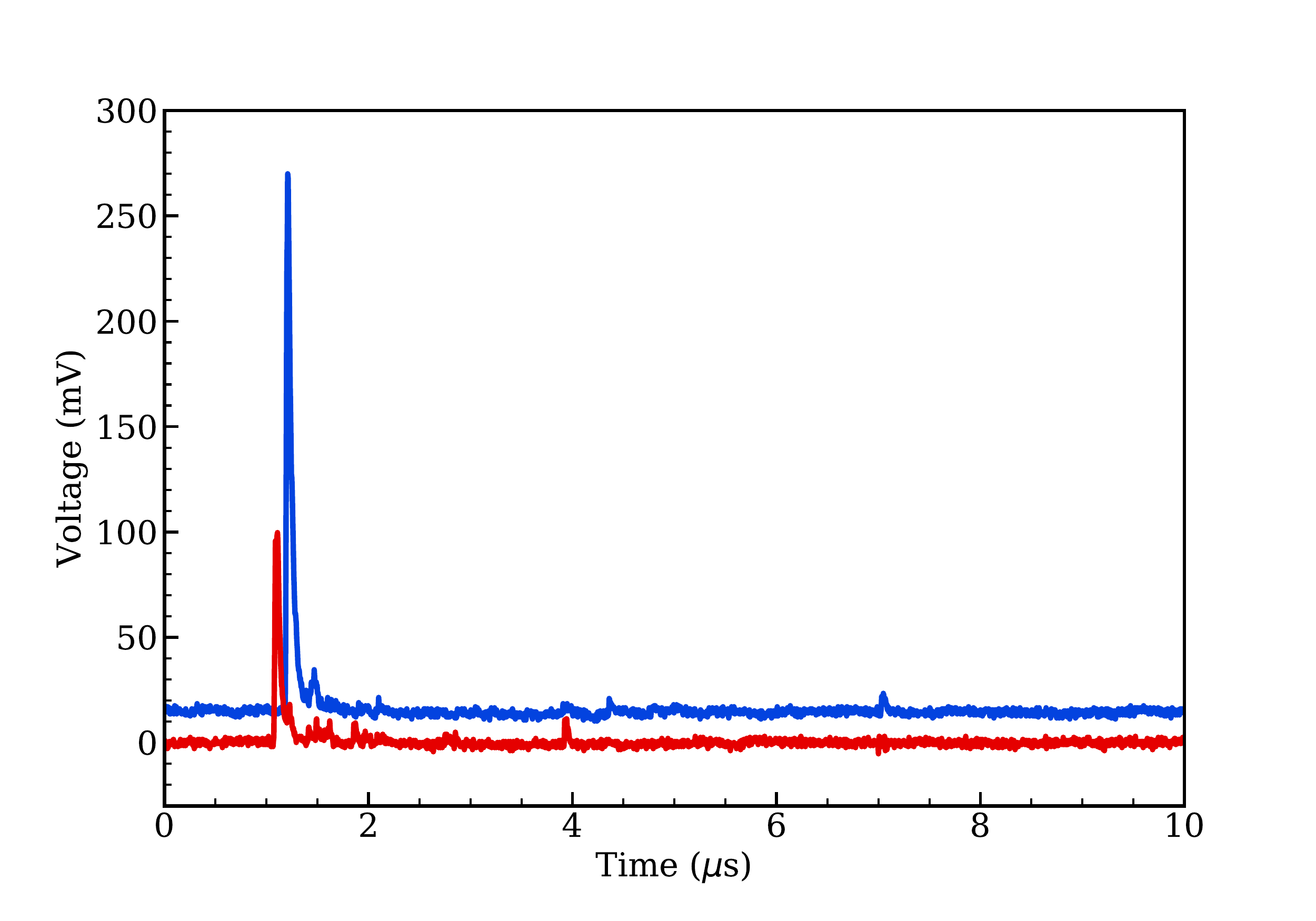}
\caption{\label{fig:sample_event} The summed output of the helium detector PMT channels from two sample events passing the analysis cuts. The ER event (blue), offset by +15 mV, was selected from the dataset corresponding to a recoil energy of 154 keV$_\mathrm{ee}$ and the NR event (red) to 142 keV$_\mathrm{nr}$. Both events contain small single photoelectron (SPE) pulses following the large prompt pulse.}
\end{figure}

The signals from the helium detector and far side detector PMTs (LS and NaI) were amplified and fed into a discriminator. The hardware trigger consisted of a two-fold coincidence requirement within a window of 150 ns among the helium detector PMTs for both DD and $^{137}$Cs data. For DD data, an extra coincidence between the LHe detector and the LS detector signal was required, also within a window of 150~ns. Each triggered event was recorded with a CAEN V1720 digitizer with 250 MHz sampling frequency and consisted of samples amounting to 1 $\mu$s pre-trigger and 31.7 $\mu$s post-trigger length. All six helium detector channels and the tagging detector channels were saved for each trigger. Summed outputs of the helium detector PMTs are shown for two typical events in Fig.~\ref{fig:sample_event}.

\begin{figure}[ht]
\includegraphics[width=\linewidth]{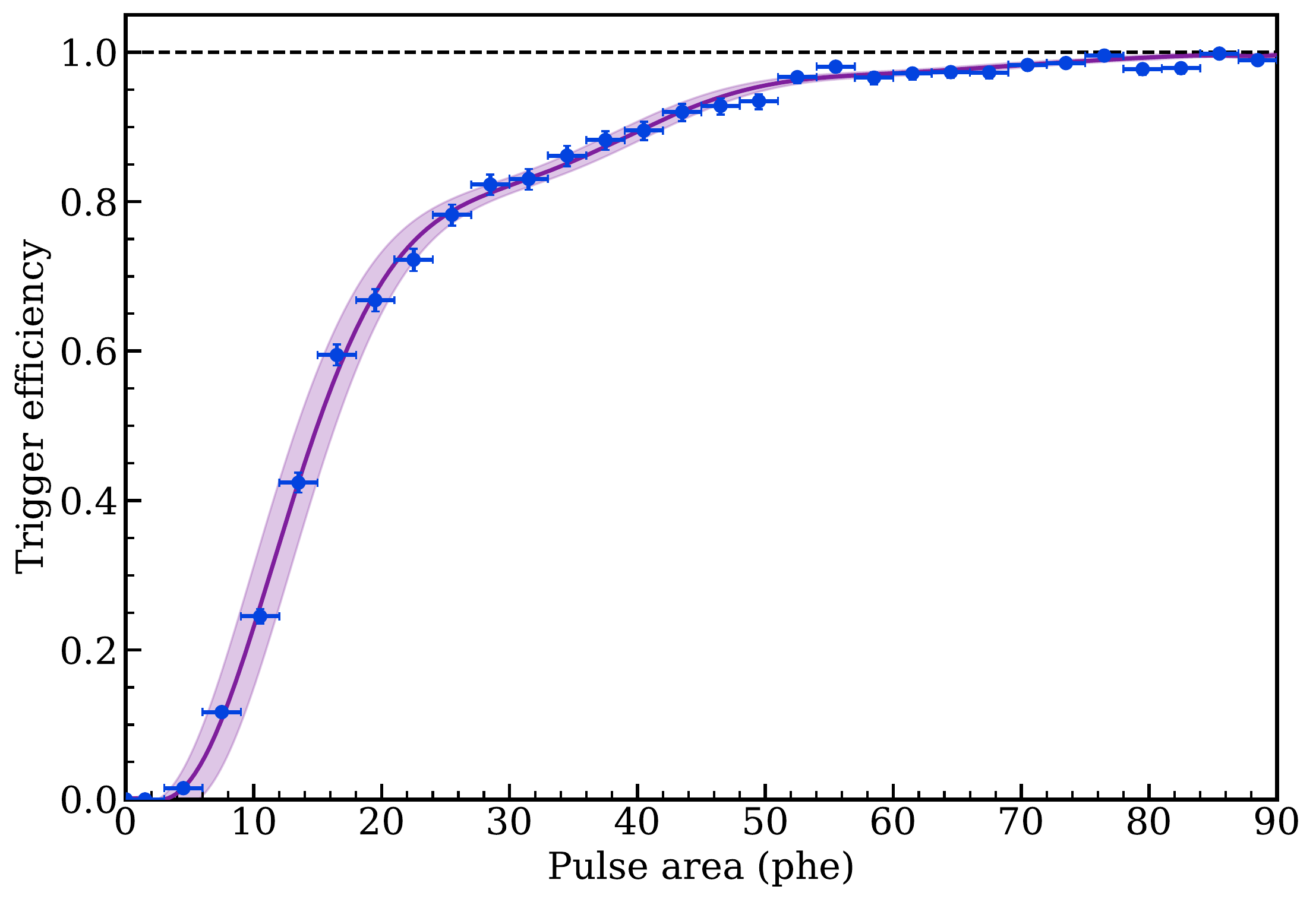}
\caption{\label{fig:efficiency_curve} The measured trigger efficiency from the combined effects of hardware triggering and software event selection as a function of signal size. A histogram of dedicated trigger efficiency data (blue) was interpolated with an 11th degree polynomial fit (purple), shown with the 1-$\sigma$ confidence band, which was derived using the horizontal and vertical error bars treated as 1-$\sigma$ errors.}
\end{figure}

The timing for the helium detector PMTs and the far side detector PMTs was synchronized by the back-to-back 511-keV gamma rays from a $^{22}$Na source. A study of the trigger efficiency for the helium detector channels was also performed with the same $^{22}$Na source, which was placed between the helium detector and a NaI detector. In this trigger efficiency study, the NaI detector signal was used to trigger the data acquisition, and the signals of the helium detector PMTs and the output from the discriminator channel linked to the helium detector PMTs were recorded. Additionally, our software analysis chain required the pulse finder to identify pulses in two channels as a data quality control. We calculated the combined pulse area from the PMTs normalized to the individual PMT gain (see Sec.~\ref{sssec:SPE_size}), and calculated the combined efficiency of triggering the discriminator and passing the analysis-level coincidence requirement as a function of pulse area. This trigger efficiency as a function of normalized pulse area is shown in Fig.~\ref{fig:efficiency_curve}. The trigger efficiency is low for fairly large signal sizes, which is postulated to be caused by the relatively low gain in the PMT signal and the fluctuation of baseline. However, most of the datasets in this study have pulse areas larger than those with low trigger efficiency, with the exception being the 53.2 ~keV$_\mathrm{nr}$ dataset.

\section{Monte Carlo simulations}
\label{sec:simulations}
To account for systematic effects due to the source spectrum, finite detector geometries and multiple scatters, the expected recoil energy spectra in the liquid helium active volume were simulated with Geant4 \cite{Agostinelli:03nimpra, Allison:IEEE, Allison:16nimpra} version~10.5, using the reference \textsc{Shielding} physics list \cite{G4_physics}. The $^{137}$Cs source was implemented as isotropic 661.7-keV gamma rays originating from the Mylar/Kapton source packaging~(Eckert \& Ziegler Type M). The angle-energy relation of the DD generator neutron production was taken from Ref.~\cite{LISKIEN1973569} and modeled as a 4th order polynomial in the simulation. 

\begin{figure}[ht]
\includegraphics[width=\linewidth]{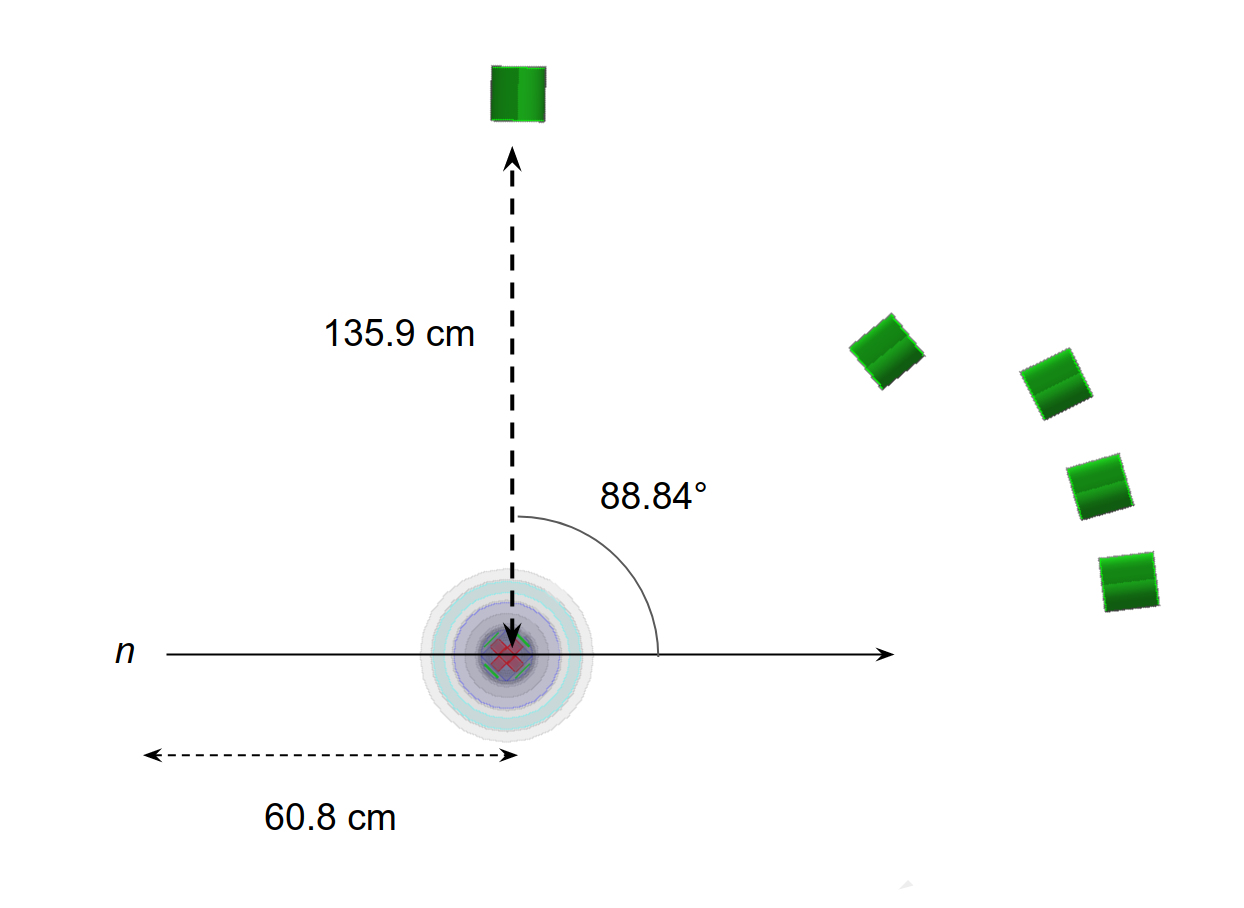}
\caption{An example simulation setup for neutron scattering. In this configuration, five far side backing detectors were simulated simultaneously for higher computational efficiency. The DD neutron source is located at the lower left corner, 60.8~cm away from the liquid helium target. In the case of Compton scattering (not shown), an isotropic gamma ray source was placed 24.89~cm from the liquid helium target along the same beam line as the neutrons, and the far side detectors were changed to NaI detectors.}
\label{fig:sim_setup}
\end{figure}

Multiple far side backing detectors corresponding to different scattering angles were placed in a single simulation run to increase computational efficiency. An example run geometry is shown in Fig.~\ref{fig:sim_setup}. When neutrons or gamma rays scattered between different far side detectors, only the first interaction was used in the subsequent analysis. During the processing, electronic recoil and nuclear recoil energies were tracked separately so that the correct signal scale could be applied in the analysis, in case a simulated event consisted of both types of interactions. 

\section{Data analysis}
\label{sec:analysis}
The experimental data were analyzed for the total signal size and the delayed scintillation behavior as a function of recoil type and energy. The signal size was found by converting the raw area of pulses to units of photoelectrons (phe) by dividing by the measured single photoelectron (SPE) size in each PMT channel, while the delayed scintillation analysis considered the temporal distribution of SPEs after a prompt window of 400~ns.

\subsection{SPE size calibration}
\label{sssec:SPE_size}
After the prompt light produced by an energy deposit, the following microseconds of an event show an elevated rate of SPEs due to delayed scintillation; some SPEs late in an event window are visible in Fig.~\ref{fig:sample_event}. This delayed signal is interesting in its own right and will be discussed further in Sec.~\ref{ssec:delayed_light_yield}. For the total signal analysis, this delayed scintillation provided a convenient source of SPEs for calibration. 

\begin{figure}[ht]
\includegraphics[width=\linewidth]{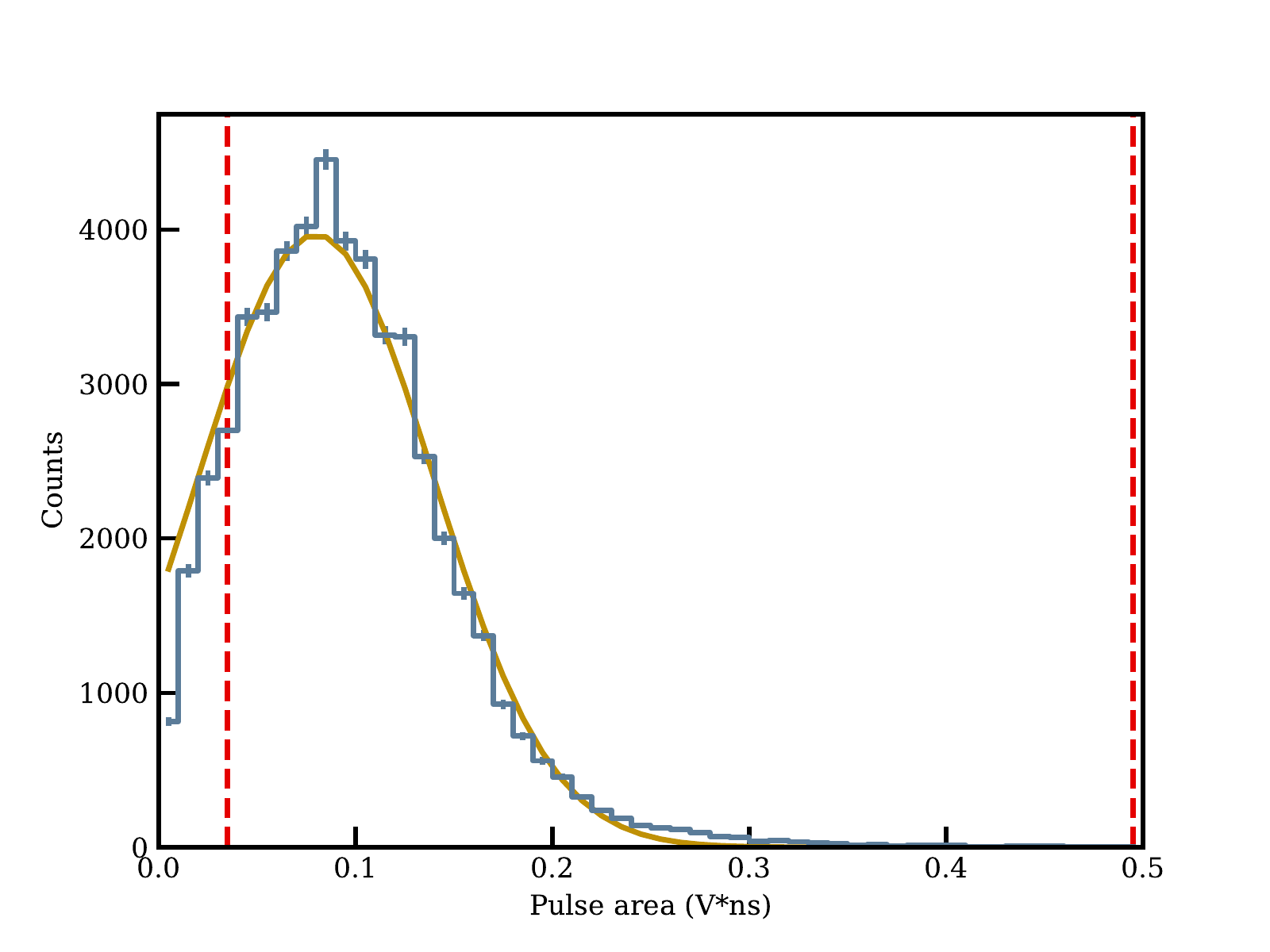}
\caption{\label{fig:spe_spectrum} Example spectrum of delayed scintillation pulses used to calibrate SPE size for a single dataset. The observed spectrum (blue) is fit to a Gaussian (yellow-orange) between the dashed red lines. Below the lower dashed line, a noise pedestal is visible in a minority ($\sim$15\%) of datasets.}
\end{figure}

To determine the SPE size, pulses were selected that arrived more than 476~ns after the nominal trigger time. The areas of selected pulses in each data taking run were histogrammed and fit to a Gaussian. An example fit is shown in Fig.~\ref{fig:spe_spectrum}. Delayed scintillation was a particularly valuable SPE source since it was observable in each data taking run, allowing changes in gain to be monitored over the course of the experiment. Each data taking run was calibrated based on the SPE sizes observed in that same set of events. A subset of the SPE sizes were checked against in-situ LED calibrations which yielded similar results.

The pulse finder removed pulses below a fixed area to avoid identifying noise as pulses. Inefficiency in finding SPEs was estimated as the fraction of the Gaussian fit to the SPE area distribution that fell below this threshold.  Across all of the PMT channels and datasets included in the analysis, the average SPE finding efficiency was 72\%. 

Of the six PMTs in the helium detector, calibration confirmed four to be usable in most data taking runs after the detector was cooled down. One channel appeared disconnected, while another demonstrated poor gain such that SPEs were not readily distinguished from noise. There were two ER datasets, about 7\% of the total ER data taking time, in which a single additional PMT was dropped from the analysis due to the fitted SPE size falling close to the baseline noise.

\subsection{Total signal size}

Each dataset consists of data taken with either the DD generator or the $^{137}$Cs source scattering into a particular recoil angle. The scintillation signal size for individual events in each dataset was determined using the SPE size calibration described in the previous section. The trailing pulse area in the event acquisition was corrected by the SPE finding efficiency for that dataset and then added to the prompt area for a total scintillation signal size in the event.

The datasets were reduced by a series of cuts based on pulse timing and the far side detector response to select for events which scatter exactly once in the liquid helium target while disfavoring multiple scatter events and those formed by accidental coincidences in the helium and tagging detector volumes. Finally, Monte Carlo (MC) simulation spectra were fitted to each dataset independently by floating a mean signal scaling parameter, a parameter for the energy resolution of the liquid helium response, and an overall scale factor of the distribution.

\subsubsection{ER data selection cuts} \label{sssec:ER_cuts}
The two tagging detectors in the ER configurations provided information about timing and recoil energy in the NaI target. Each detector was calibrated independently for both timing and energy resolution.

\emph{Timing cut}---As mentioned in Section~\ref{ssec:data_acq_and_trigger_eff}, we synchronized the helium detector and tagging detector timing with the back-to-back gamma rays from $^{22}$Na decays. For each tagging detector, we fit a Gaussian to the distribution of pulse time differences between that detector and the helium detector to calibrate the time resolution of the experimental configuration. We found a resolution of 4~ns (sigma) for both NaI detectors, comparable to the 1-2~ns it takes the scattered gamma ray to travel between the liquid helium and NaI volumes in the Compton scattering data taking configurations. 

For the timing cut in the Compton scattering analysis, we used a relatively wide coincidence window of $\pm$20 ns since we found that a tighter definition had no strong effects when used in conjunction with the NaI energy deposit cut described below. The same timing cut was applied to the MC spectra after applying Gaussian smearing with the measured 4~ns resolution. We estimated the systematic uncertainty associated with this cut by rerunning the analysis with modified values of $\pm$16~ns and $\pm$24~ns and found variations at 1\% or less for all datasets except 98.9 keVee, which had a 5\% variation in the best fit value.

\emph{NaI energy deposition cut}---Gamma rays which Compton scatter once in the helium target have an outgoing energy
\begin{equation}
\label{eq:Compton_scattering}
E_\gamma' = \frac{E_\gamma}{1+(E_\gamma/m_ec^2)(1-\mathrm{cos}~\theta)},
\end{equation}
where $E_\gamma$ is the ingoing energy of 661.7 keV, $m_e$ is the mass of the electron, $c$ is the speed of light, and $\theta$ is the lab frame scattering angle. $E_\gamma'$ ranges from 476 to 625 keV for the recoil energies in this study. NaI detector scintillation can be used to reconstruct the deposited energy to a high degree of accuracy, providing an additional data selection cut by looking for events which match the expected remaining gamma ray energy in the NaI detector.

We used the decay gamma rays from $^{22}$Na, $^{57}$Co, $^{133}$Ba, and $^{137}$Cs to calibrate the NaI detector response from 122 to 1275 keV. First, the mean response at each gamma ray energy was found by fitting a Gaussian to the photoabsorption peak. Then, we used a linear fit to determine the detector response as a function of energy. Similarly, the behavior of the energy resolution was estimated by fitting a function of the form
\begin{equation}
\sigma_r(E) = c_1 E + c_2 E^{1/2} + c_3,
\end{equation}
to the standard deviations obtained from the Gaussian fits, where the $c_i$ are fit coefficients for each term. Our measured single standard deviation resolution for gamma ray energies in the range expected of $E_\gamma'$ was 3\%-4\%.

\begin{figure}[ht]
\includegraphics[width=\linewidth]{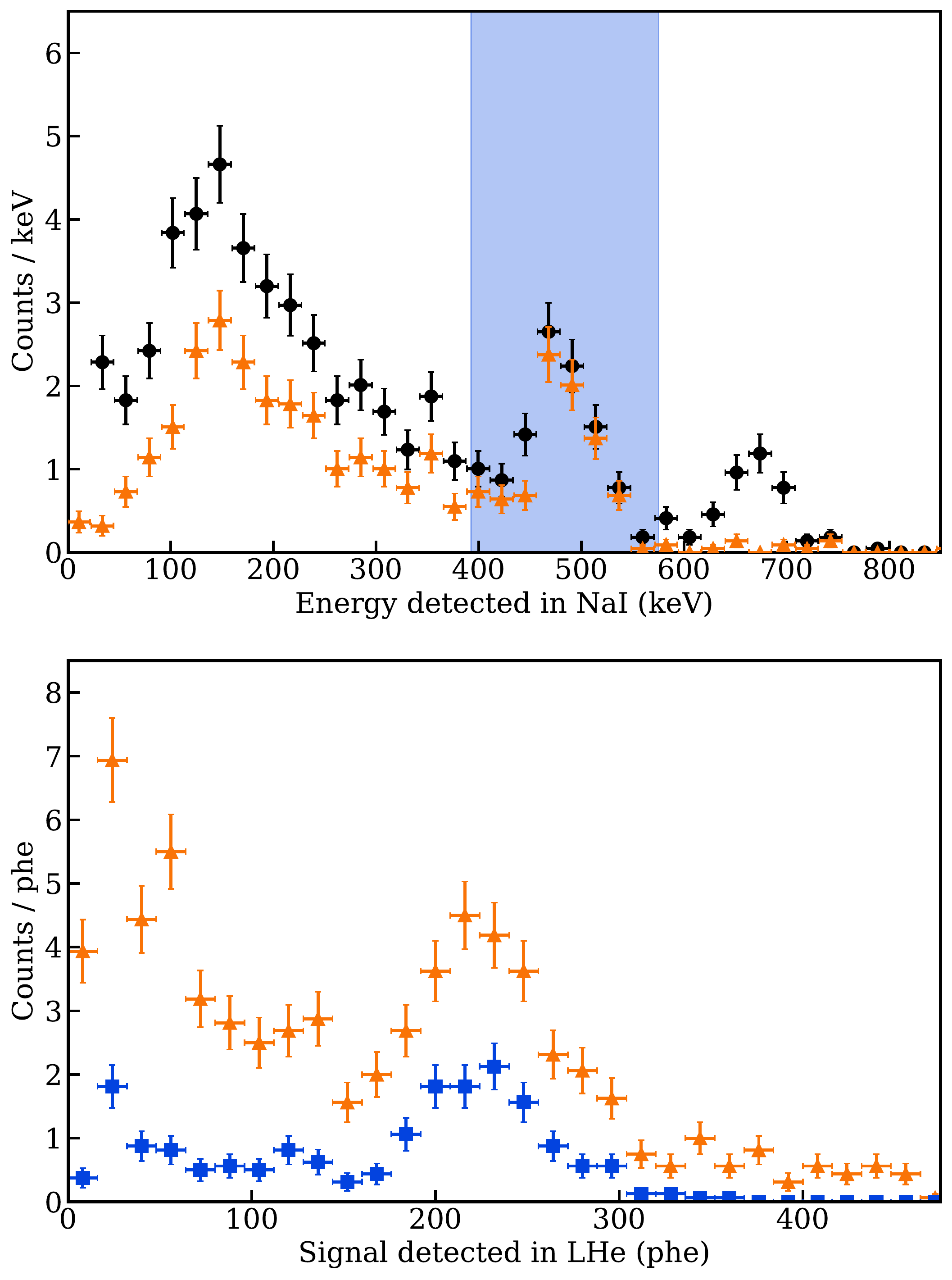}
\caption{\label{fig:NaI_cut} The 185~keV$_{\mathrm{ee}}$ dataset as an example of the NaI cut applied to all of the Compton scattering datasets. The response in the NaI detector (top) was used to select events more likely to match the energy $E_\gamma'$ expected from Eq.~(\ref{eq:Compton_scattering}). All events before cuts are shown in the black circles, where the photoabsorption peak for 661.7 keV $^{137}$Cs gamma rays is a visible feature. It mostly vanishes as a result of the timing cut after which the orange triangle events remain. Still present in the orange spectrum is a photoabsorption peak corresponding to $E_\gamma' = 476$~keV. Events which fall inside of the blue band are accepted for the analysis (the definition of the band is described in the text). The effect of the NaI energy cut on the helium scintillation spectrum (bottom), where the orange triangle events correspond to those in the top panel and the blue square events are those which also pass the NaI energy cut, is a more prominent peak attributed to single scatter events.}
\end{figure}

For each dataset in the analysis, events were cut according to the signal response in the NaI detector. Events were accepted if the response in the NaI detector was within $E_\gamma' \pm 5\sigma_r(E_\gamma')$. The effect of the cut is demonstrated in the bottom panel of Fig.~\ref{fig:NaI_cut}, where the orange population consists of events with only the timing cut described above and the blue spectrum represents those also passing the NaI energy cut. While this cut removes some events in the single scatter peak visible in the liquid helium signal spectrum, selecting events consistent with an energy deposit of $E_\gamma'$ in the NaI boosted the ratio of events in the liquid helium single scatter peak relative to its side bands. The NaI energy deposit cut was replicated in the MC events by applying the measured energy resolutions of the two tagging detectors. The distinct calibrations of the two tagging detectors were weighted according to the number of events collected with each detector in a given dataset. The systematic uncertainty resulting from this cut definition was estimated by rerunning the analysis with $\pm 4\sigma_r$ and $\pm 6\sigma_r$ as the selection window, resulting in variations at the 0.5-3\% level.

\subsubsection{NR data selection cuts} \label{sssec:NR_cuts}

NR datasets were reduced using pulse shape discrimination and timing cuts enabled by the LS tagging detector. 

\begin{figure*}[ht]
\includegraphics[width=\textwidth]{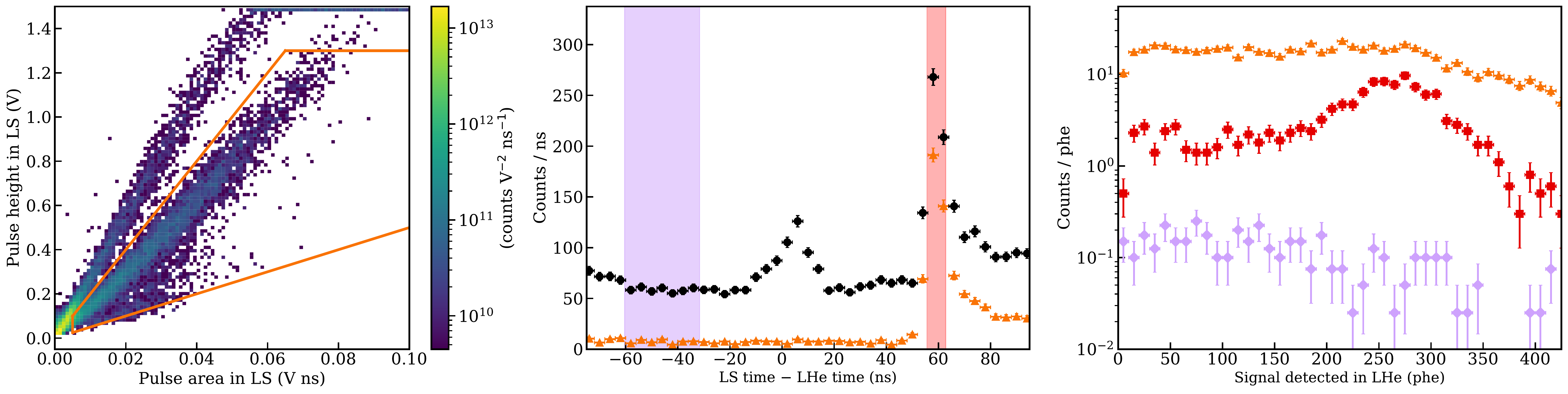}
\caption{\label{fig:LS_ToF_cuts} The 561 keV$_{\mathrm{nr}}$ dataset as an example of the cuts applied to the nuclear recoil datasets. The left panel depicts the pulse shape discrimination (PSD) of the liquid scintillator (LS) tagging detector. The upper band comes from electronic recoil events and the lower band from nuclear recoils. Events from the lower band inside of the orange lines are accepted and tagged as neutrons. The middle panel shows the effect of this cut on the event distributions in time-of-flight, defined as the liquid scintillator time minus the event time in the liquid helium. The spectrum marked by black circles consists of all events before cuts, while events plotted in orange triangles are those tagged as neutron scatters by the PSD cut. The prompt peak in the total event spectrum is consistent with scatters produced by gammas rather than neutrons as it disappears on application of the PSD cut. The peak in neutron events first arriving about 53 ns later corresponds to neutrons scattering once in the liquid helium, while the longer tail comes from multiple scatter events. Events in the red band are accepted as single scatter neutron events, and appear as red squares in the right panel. We estimated the random coincidence backgrounds with the neutron scatters in the purple band. The liquid helium scintillation spectrum of neutron-tagged events is plotted as orange triangles in the right panel. Events passing the time-of-flight cut are shown as red squares. The accidental coincidence event spectrum was rescaled according to the size of the time-of-flight acceptance window and plotted as purple diamonds.}
\end{figure*}

\emph{Pulse shape discrimination cut}---The BC-501A LS pulse shape can be used to discriminate recoil types by considering the maximum pulse height and total pulse area in the event. An example of the pulse shape discrimination (PSD) cut for the 561~keV$_{\mathrm{nr}}$ dataset is shown in the left panel of Fig.~\ref{fig:LS_ToF_cuts}; electronic recoils form a clear upper band and nuclear recoils form the lower band. Events from the lower band within the orange lines were tagged as neutron scatters in the liquid scintillator. We applied a minimum area cut of 5~V\,ns because the two bands merge at low area, reducing the discrimination power. A maximum height cut was used to eliminate events saturating the digitizing electronics. Since the PSD bands were clearly separated for large area events, the systematic uncertainty estimate for this cut was to vary the minimum area to 2.5~V\,ns and 10~V\,ns which resulted in at most 1\%-level variation in the final signal scaling parameter values.

The effect of the PSD cut is shown in the middle panel of Fig.~\ref{fig:LS_ToF_cuts}, which shows spectra of the time-of-flight, defined as the event time in the LS minus the event time in the liquid helium. All of the events in the dataset are shown in the black spectrum, while events tagged as neutrons by the PSD cut are plotted in orange. A prompt peak near a time-of-flight of 0~ns from gamma scatters is clearly visible in the events before the application of the PSD cut, and a peak in the tagged neutron events is apparent around the expected time-of-flight for the experimental configuration.

As with other cuts, this cut was replicated in the MC events. However, the effect of this cut on the MC events was minuscule because the majority of the events in the electronic recoil band of the LS discrimination plot were from accidental coincidences with background gammas, which were not simulated.

\emph{Time-of-flight cut}---Neutrons which elastically scatter once in the helium target into a known angle deposit a well-defined energy
\begin{multline}
    \label{eq:neutron_elastic_scattering}
    E_r = \frac{2 m_n E_n}{(m_n + m_\mathrm{He})^2}\left[\vphantom{\sqrt{\sin^2(\theta)}}m_n \sin^2(\theta) + m_\mathrm{He} \right. \\
    \left. - \cos(\theta)\sqrt{m^2_\mathrm{He} - m_n^2\sin^2(\theta)} \right],
\end{multline}
where $E_r$ is recoil energy of the helium atom, $m_n$ is the mass of the neutron, $E_n$ is the initial energy of the neutron, $m_\mathrm{He}$ is the mass of the helium atom, and $\theta$ is the scattering angle of the neutron in the lab frame relative to its initial direction. The time it takes a neutron to travel between the liquid helium volume and the tagging detector is well-determined by its energy, so an additional constraint on good single scatter events is the time measured between events in the LHe target and the LS tagging detector.

The time resolution of the setup with the LS tagging detector was measured with a $^{22}$Na source in the same way as for the NaI detectors, resulting in a time resolution of $\sigma_t$ = 2.4 ns. The middle panel of Fig.~\ref{fig:LS_ToF_cuts} shows a tagged neutron peak consistent with the expected 53 ns time-of-flight for single scatter neutrons in this experimental geometry. A time-of-flight cut was used to select these single scatter events and reject neutrons scattering multiple times before reaching the tagging detector, which do not necessarily deposit the energy described by Eq.~(\ref{eq:neutron_elastic_scattering}). Events within -1/+2$\sigma_t$ around the peak in the spectrum were accepted, where the bounds were optimized by examining the time-of-flight behavior in MC events and also by maximizing the liquid helium scintillation peaks relative to their side bands in experimental data. As with other cuts in the analysis, we modified these bounds to -2/+2$\sigma_t$ and -1/+3$\sigma_t$ as a systematic uncertainty estimate and found variations between 1-2\%.
 
The time-of-flight acceptance region is the red band in the middle panel of Fig.~\ref{fig:LS_ToF_cuts}, and the events that pass both cuts are shown in red in the right panel. The red distribution reveals a well-defined scintillation signal size peak associated with single scatters in the LHe. There is a flat distribution of nuclear recoil events in the time-of-flight spectrum arising from accidental coincidences in the two detector volumes. The contribution of such events to the signal region was estimated from the purple region in the middle panel of Fig.~\ref{fig:LS_ToF_cuts}, scaled to the width of the time-of-flight acceptance window. This scaled spectrum is shown in purple in the right panel of Fig.~\ref{fig:LS_ToF_cuts}, where it is clearly subdominant to the main signal events in red.

\subsubsection{Fitting procedure} \label{sssec:fit_procedure}

MC events were first converted from units of energy to signal size in number of photons. The number of photons $S$ was determined by 
\begin{equation}
    \label{eq:light_yield_definition}
    S = Y_{xx} E_{xx},
\end{equation}
where $Y_{xx}$ are the signal scaling parameters with units phe/keV and $E_{xx}$ are the simulated deposit energies of each type in units of keV. 

Next, the MC events were weighted according to the selection cuts and smeared using a Gaussian energy resolution function
\begin{equation}
    f(x, S, A) = \frac{1}{\sqrt{2\pi A^2S}}\exp\left[\frac{-(x-S)^2}{2A^2S}\right],
\end{equation}
where $x$ is the smeared signal in phe, $S$ is the signal from the MC event given by Eq.~(\ref{eq:light_yield_definition}), and $A$ is a resolution factor with units $\sqrt{\mathrm{phe}}$. The fitted trigger efficiency curve from Fig.~\ref{fig:efficiency_curve} was applied to the smeared spectrum by integrating the function over each bin in the smeared spectrum and rescaling the bin by the result. It is possible, though unlikely, for DD neutron events to induce ERs in the liquid helium target through inelastic scatters and neutron capture on materials around the target. After the time of flight selection, fewer than 0.1\% of MC neutron events had an ER energy component. Thus, we neglected the ER energy component in our treatment of the NR MC. 

The smeared MC spectra were each separately fit to histograms of the experimental data passing selection cuts with three floating parameters: $Y_{ER/NR}$, $A$, and a third parameter determining the overall height of the MC spectrum. Fitting was performed by minimizing the $\chi^2$ test statistic
\begin{equation}
\chi^2 = \sum_{i=1}^{N}\frac{(n_i-\nu_i)^2}{\nu_i},
\label{eq:chi2}
\end{equation}
where $n_i$ is the observed number of events in each bin, $\nu_i$ is the expected number of events in each bin from the smeared MC spectrum, and $i$ runs over the $N$ bins in the fit region. 

The fit region was determined iteratively, first by hand selecting the boundaries and finding the best fit parameters. Then,  those parameters were used to smear the distribution of the single scatter events of interest in the MC, and the iterated fit region was defined as $\pm 2 \sigma$ about the mean of a Gaussian fit to those events. The final signal scaling parameter results were those resulting from rerunning the fitting procedure using the iterated fit region.

\begin{figure}[ht]
\includegraphics[width=\linewidth]{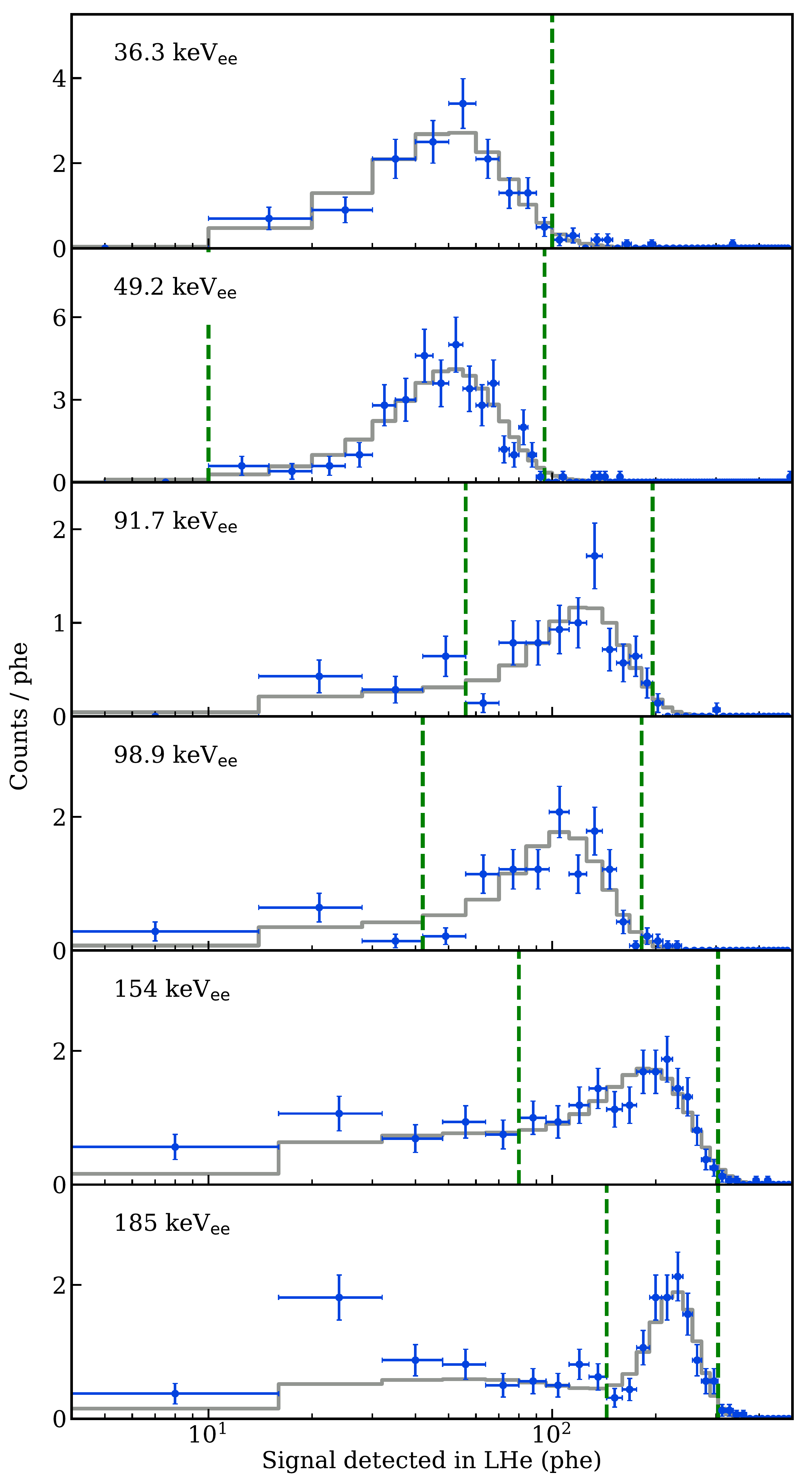}
\caption{\label{fig:ER_fits} Histograms of experimental data (blue) and fitted MC (gray) for each Compton scattering recoil energy. The x-axis showing the signal size in the liquid helium is the same for each panel. The fit region for each fit is between the vertical green dashed lines.}
\end{figure}

\begin{figure}[ht]
\includegraphics[width=\linewidth]{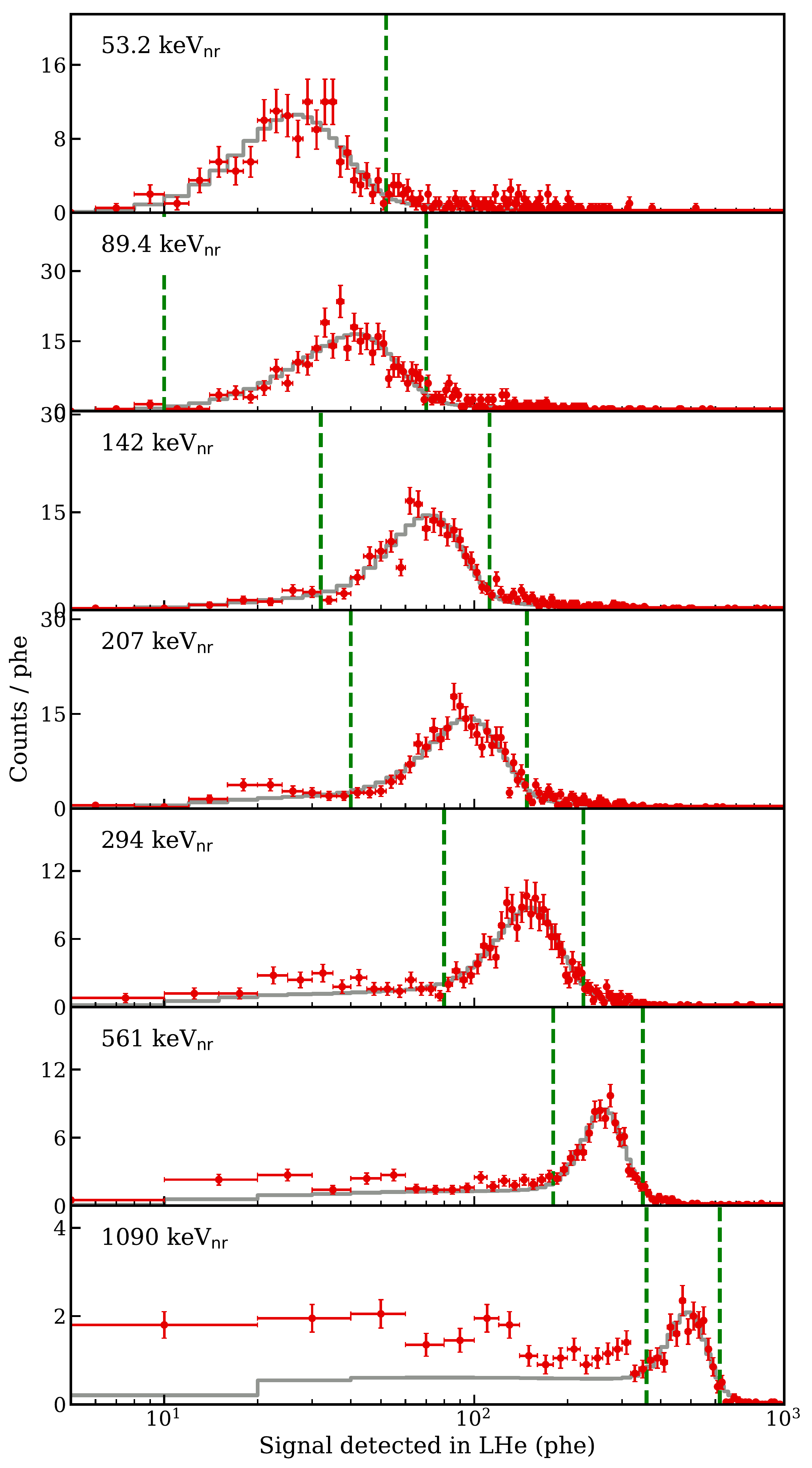}
\caption{\label{fig:NR_fits} Histograms of experimental data (red) and fitted MC (gray) for each DD neutron scattering recoil energy. The x-axis showing the signal size in the liquid helium is the same for each panel. The fit region for each fit is between the vertical green dashed lines, and the lower bound for the 53.2 keV$_{\mathrm{nr}}$ fit extends to a signal size of 0 phe.}
\end{figure}

Table~\ref{tab:errors} lists the six fitted ER and seven fitted NR signal scaling parameters, along with statistical errors from the fits and systematic uncertainties estimated as described below. Figures~\ref{fig:ER_fits} and \ref{fig:NR_fits} show histograms of the detected scintillation response in the helium target and the corresponding best fit MC spectra for the ER and NR datasets, respectively.

\setlength{\tabcolsep}{12pt}
\renewcommand{\arraystretch}{1.5}
\begin{table*}[ht]
    \centering
    \caption{Fitting results for the ER and NR datasets. Uncertainties for the recoil angles and energies were computed using the effect of position error uncertainties from the target helium volume and the tagging detector on the mean recoil energy. Uncertainties in the signal scaling parameter, $Y$, are separated into systematic uncertainty (consisting of the effects from the data selection cuts, the uncertainty due to the trigger efficiency, the recoil energy uncertainty, and the statistical uncertainty from the SPE size fits) and statistical uncertainty from the $\chi^2$ minimization fitting. We also report the fitted resolution parameter $A$ and its combined statistical and systematic uncertainty. Finally, we report the minimized $\chi^2$ and number of degrees of freedom for each of the datasets.}
    \begin{tabular}{c | c c | c c c | c | c}
    \hline\hline
    & $\theta$ (degree) & $E_r$ (keV) & $Y$ (phe/keV) & $\delta Y_\mathrm{sys}$ & $\delta Y_\mathrm{stat}$ & $A$ $(\sqrt{\mathrm{phe}})$ & $\chi^2 / \mathrm{n.d.f.}$\\
    \hline
    \multirow{6}{*}{ER} & $17.2 \pm 0.4$ & $36.3 \pm 1.5$ & 1.43 & $^{+0.10}_{-0.09}$ &                     $^{+0.11}_{-0.09}$ & $2.40^{+0.83}_{-0.67}$ & 6.1 / 7\\
                        & $20.3 \pm 0.4$ & $49.2 \pm 1.8$ & 1.08 & $^{+0.05}_{-0.05}$ & $^{+0.05}_{-0.05}$ & $1.56^{+0.66}_{-0.53}$ & 16.8 / 14\\
                        & $28.9 \pm 0.4$ & $91.7 \pm 2.2$ & 1.44 & $^{+0.06}_{-0.08}$ & $^{+0.09}_{-0.08}$ & $2.46^{+1.24}_{-0.65}$ & 10.2 / 7\\
                        & $30.2 \pm 0.4$ & $98.9 \pm 2.3$ & 1.15 & $^{+0.08}_{-0.08}$ & $^{+0.05}_{-0.05}$ & $2.39^{+0.94}_{-0.51}$ & 15.5 / 7\\
                        & $40.0 \pm 0.4$ & $154.0 \pm 2.3$ & 1.33 & $^{+0.08}_{-0.04}$ & $^{+0.05}_{-0.05}$ & $3.57^{+0.93}_{-0.75}$ & 7.9 / 11\\
                        & $45.6 \pm 0.4$ & $185.0 \pm 2.4$ & 1.27 & $^{+0.04}_{-0.04}$ & $^{+0.03}_{-0.03}$ & $1.74^{+0.43}_{-0.45}$ & 8.3 / 7\\
    \hline
    \multirow{7}{*}{NR} & $15.9 \pm 0.2$ & $53.2 \pm 1.1$ & 0.48 & $^{+0.02}_{-0.05}$ &                                      $^{+0.03}_{-0.03}$ & $2.00^{+0.65}_{-0.17}$ & 20.1 / 23\\
                        & $20.7 \pm 0.2$ & $89.4 \pm 1.4$ & 0.45 & $^{+0.01}_{-0.01}$ & $^{+0.02}_{-0.01}$ & $2.09^{+0.29}_{-0.39}$ & 35.0 / 27\\
                        & $26.3 \pm 0.2$ & $142.0 \pm 1.7$ & 0.50 & $^{+0.01}_{-0.01}$ & $^{+0.01}_{-0.01}$ & $2.06^{+0.44}_{-0.20}$ & 25.0 / 17\\
                        & $31.9 \pm 0.2$ & $207.0 \pm 2.1$ & 0.47 & $^{+0.01}_{-0.01}$ & $^{+0.01}_{-0.01}$ & $2.44^{+0.25}_{-0.20}$ & 39.7 / 24\\
                        & $38.6 \pm 0.2$ & $294.0 \pm 2.4$ & 0.52 & $^{+0.01}_{-0.02}$ & $^{+0.01}_{-0.01}$ & $2.66^{+0.30}_{-0.43}$ & 21.9 / 26\\
                        & $55.8 \pm 0.2$ & $561.0 \pm 2.8$ & 0.47 & $^{+0.01}_{-0.01}$ & $^{+0.01}_{-0.01}$ & $2.22^{+0.41}_{-0.20}$ & 12.2 / 14\\
                        & $87.6 \pm 0.2$ & $1090.0 \pm 2.6$ & 0.45 & $^{+0.02}_{-0.01}$ & $^{+0.01}_{-0.01}$ & $2.82^{+0.64}_{-0.53}$ & 10.3 / 10\\
    \hline\hline
    \end{tabular}
    \label{tab:errors}
\end{table*}

Within each dataset fit, the application of the energy resolution assumed the resolution scales with the square root of the signal size. However, the resolution factor was allowed to float across the dataset fits to make fewer assumptions about the underlying physics of the signal generation. Still, the best-fit values of the resolution factor were self-consistent across all of the datasets. Fixing the resolution factor to the mean value obtained from these fits and re-fitting the data did not yield significant differences in the final values of the signal scaling parameters or degrade the overall goodness of fit of the results.

An excess of events at small signal size for datasets corresponding to larger recoil energies was present in both the ER and NR data. One possible explanation might be varying light collection efficiency over the liquid helium target volume, but we did not find any evidence that such regions exist in the data. Most events in the final datasets were composed of roughly equal fractions of photons seen in each PMT. It is also unlikely to be the result of degraded recoil tracks, since the recoil track length in 1.75 K liquid helium for 1 MeV recoil alphas is about 40 $\mu$m and for 200 keV recoil electrons about 3 mm \cite{NIST_STAR}. This explanation is especially unlikely to account for the excess in the NR data, where the effect is most apparent. Ultimately, the definition of the fit region around the single scatter peak excluded these low energy excesses from the fits.

\subsubsection{Systematic uncertainties}

Systematic uncertainties associated with the data selection cuts were estimated by varying those cuts and rerunning the fit for each dataset, as previously discussed in the descriptions of the cuts. The systematic uncertainty associated with each cut was estimated as the difference between the best fit value from the main analysis and the fit with the varied cut parameters. These cut-based systematic uncertainties each contributed at the percent level for all of the datasets in the analysis.

The systematic uncertainties due to the fit region definition and the application of threshold efficiency curve were also estimated in this way. The helium signal fit region for each dataset was varied from $\pm 2 \sigma$ of the Gaussian fit on the MC single scatter distribution to $\pm 1.5 \sigma$ and $\pm 2.5 \sigma$. As with the cut-based systematic uncertainties, the fit region definition contributed at the 1-2\% level for all of the datasets, except for a 4\% difference in the 154 keVee dataset. Likewise, the threshold efficiency curve was replaced by the lower and upper 1-$\sigma$ bounds shown in Fig.~\ref{fig:efficiency_curve}. The effect of this change was negligible for almost all of the datasets, since the fit regions generally excluded the signal sizes for which the threshold curve strongly varied. However, the systematic uncertainty for the 53.2 keV$_\mathrm{nr}$ fit, the smallest signal size across all of the datasets, was dominated by the uncertainty in the threshold efficiency curve with the best fit value varying -8\% and +3\% for the lower and upper threshold efficiency variations.

Uncertainties on the positions of the target and tagging detectors also contributed some uncertainty to the energies probed by each dataset. We estimated the size of this effect with a toy Monte Carlo approach by varying the positions of the detectors according to the 1~mm position uncertainty and calculating the nominal recoil energy for the modified configurations. A Gaussian was fit to the distribution of nominal recoil energies to translate the effect of the position uncertainty to an uncertainty in the recoil energy assigned to each dataset. This error is listed in Table \ref{tab:errors}, plotted as horizontal error bars in Figs.~\ref{fig:ER_LY_plot} and \ref{fig:relative_light_yield}, and folded into the systematic uncertainties on the measured signal scaling parameter values. This uncertainty in the recoil energy is distinct from the range of recoil energies sampled by the experimental geometry due to the finite sizes of the detector elements, which is accounted for by the simulation geometry in Geant4.

The uncertainty due to the SPE size calibration was estimated using the statistical error from the SPE fits for individual PMTs in each data taking run. Since the datasets consisted of events from multiple data taking runs, the signal size uncertainty for each dataset was estimated as the average of the SPE size uncertainties weighted according to the number of events from each data taking run in the dataset. This uncertainty contributed at the 2\%-3\% level for the NR datasets. It contributed a higher 3\%-5\% uncertainty in the signal scaling parameter of the ER datasets due to fewer SPE statistics for these acquisitions.

As described in Sec.~\ref{sssec:SPE_size}, the effect of the SPE finding efficiency was corrected for in the total signal size associated with each event. The uncertainty in this efficiency was determined from the Gaussian fit parameters of the SPE area distribution, and propagated to the uncertainty in the total signal size for each dataset. The uncertainty associated with this correction was comparable to other systematic errors described in this section at 1\%-5\% in the best fit signal scaling parameter.

\subsection{Delayed scintillation} \label{ssec:delayed_light_yield}

In addition to our study of the total signal, we quantified the partitioning of the scintillation in time among the prompt, exponential, and $t^{-1}$ components.

Events in this analysis were selected using the same selection cuts described for the total signal. While the prompt scintillation caused a high amplitude signal in each PMT from many photons spaced closely in time, the delayed components consisted of well separated SPE pulses (see Fig.~\ref{fig:sample_event}). Unfortunately, low PMT gains relative to noise during this experiment led to a low efficiency of finding SPEs in most channels. Therefore, we restricted the delayed scintillation analysis to pulses found in a single PMT channel that demonstrated higher gain and higher SPE efficiency.

\begin{figure}[ht]
\includegraphics[width=\linewidth]{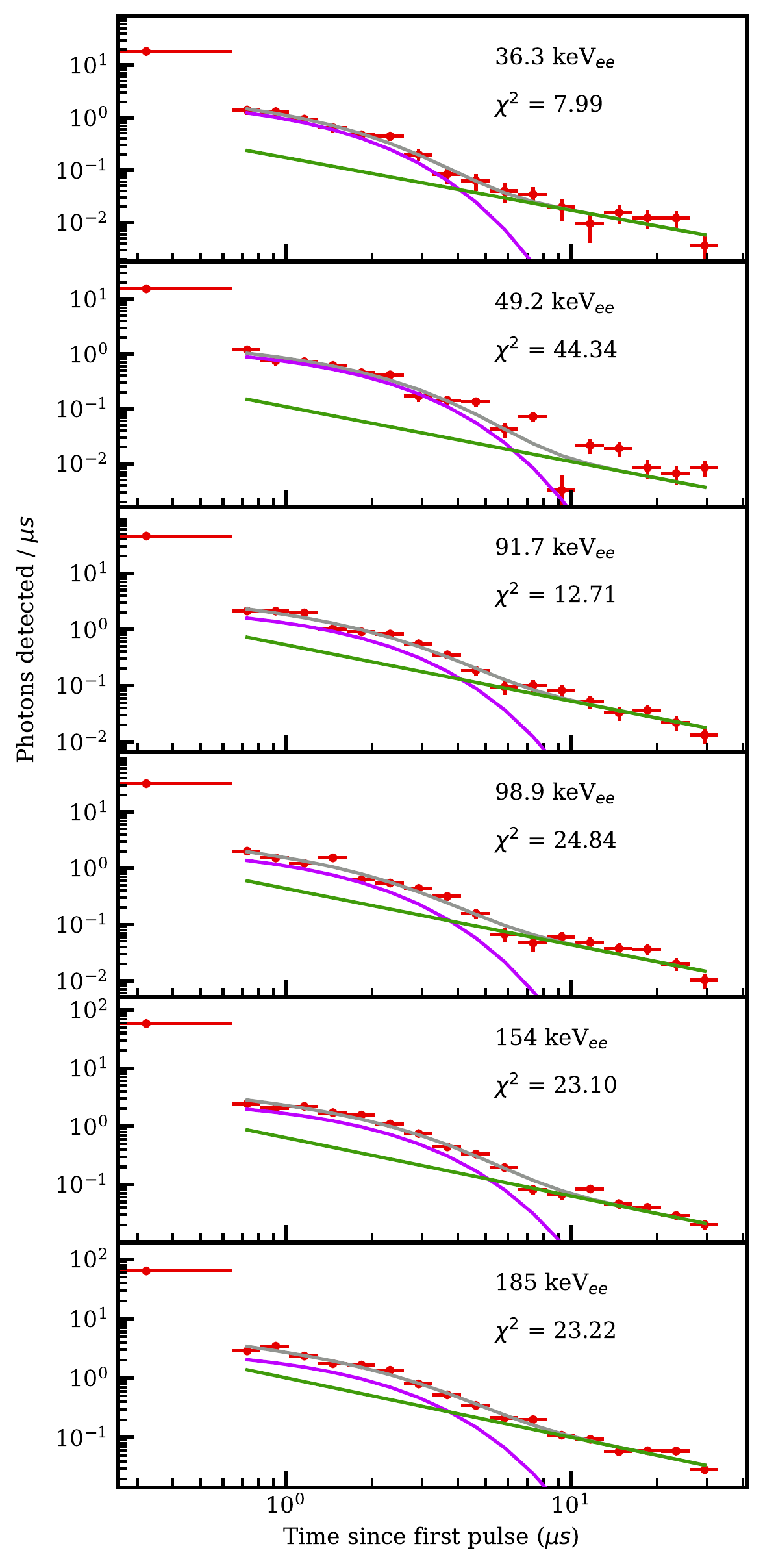}
\caption{\label{fig:er_afterpulsing_fits} ER delayed scintillation fits of exponential (purple) and 1/t (green) components to data (red). The total model is the sum of the two fitted components (gray). The $\chi^2$ value is shown for each fit; for each fit there are 14 degrees of freedom.}
\end{figure}

\begin{figure}[ht]
\includegraphics[width=\linewidth]{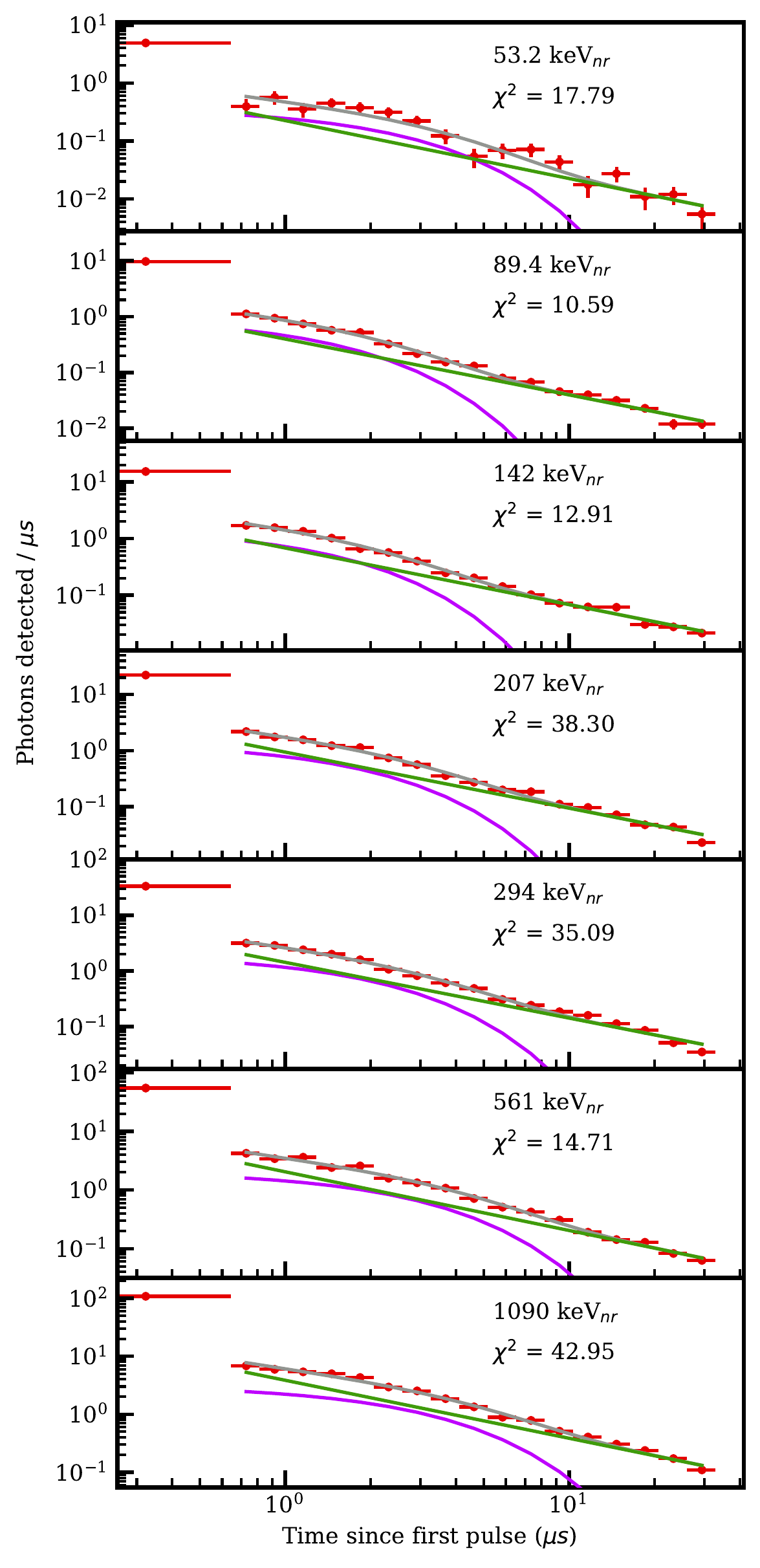}
\caption{\label{fig:nr_afterpulsing_fits} NR delayed scintillation fits of exponential (purple) and 1/t (green) components to data (red). The total model is the sum of the two fitted components (gray). The $\chi^2$ value is shown for each fit; for each fit there are 14 degrees of freedom.}
\end{figure}

For each recoil energy, the pulse times across all events were combined into a single histogram. A fit to the model 
\begin{equation}
\label{eq:delayed_scint_model}
    n_{phe}(t) = Ae^{-\lambda t}+Bt^{-1}
\end{equation}
was performed over times above 640~ns, where $t=0$ was defined as the time of the prompt pulse, and the amplitudes $A$ of the exponential component and $B$ of the $t^{-1}$ component, as well as the decay rate $\lambda$ of the exponential component were free parameters. The fits can be seen for ER data in Fig.~\ref{fig:er_afterpulsing_fits} and for NR data in Fig.~\ref{fig:nr_afterpulsing_fits}. Pulses before 400~ns were often merged with the initial pulse; analysis of smaller timescales was not pursued in this study. $\chi^2$ values for the fits are shown, and residuals from the fits do not suggest any clear modification to the time dependence for improved modeling. At timescales beyond the $\approx 32~\mu s$ event window, we expect a scintillation component due to triplet decay with a lifetime of 13~s. However, within the event window, any constant contribution to the rate was found to be negligible. This also indicates that PMT dark rate does not contribute significantly to the delayed SPE rate.

The fraction of scintillation assigned to each component was determined by integrating each of the terms in Eq.~(\ref{eq:delayed_scint_model}) from 640~ns to 32~$\mu $s, and assigning the remainder to the prompt scintillation. Along with statistical error, two sources of systematic uncertainty were included in the total error bars presented for the scintillation fraction. The uncertainty associated with the choice of the lower bound of the fit window was estimated by varying the lower bound of the fit window between 400~ns and 1.6~$\mu$s (while keeping the lower integration bound fixed at 640~ns). The fits displayed in Figs.~\ref{fig:er_afterpulsing_fits} and \ref{fig:nr_afterpulsing_fits} and central values on scintillation fractions in Figs.~\ref{fig:er_scint_partition} and \ref{fig:nr_scint_partition} were computed with a lower fit window bound of 640~ns. As described in Sec.~\ref{sssec:SPE_size}, the SPE finding efficiency was estimated as the fraction of the Gaussian fit to the SPE area distribution that fell below a fixed threshold. Uncertainty in the Gaussian fit parameters was propagated to the uncertainty in SPE finding efficiency, and then to the uncertainty in scintillation fractions. Statistical uncertainty dominated over both of these systematics for most data points.

\section{Results and discussion}
\label{sec:discussion}
Along with discussion of the results from the total and delayed scintillation, we provide a brief overview of the semiempirical model we use for comparison, which we previously described in Ref.~\cite{Hertel:2019}. The ER modeling remains largely unchanged from Ref.~\cite{Hertel:2019}, while the details described here concern the calculation of the NR energy partitioning into different signal channels. We also detail the addition to that model of an estimate of delayed scintillation from triplet-triplet quenching following Ref.~\cite{King:66}.

\subsection{Model overview}
\label{ssec:model_description}

\begin{figure}[ht]
\includegraphics[width=\linewidth]{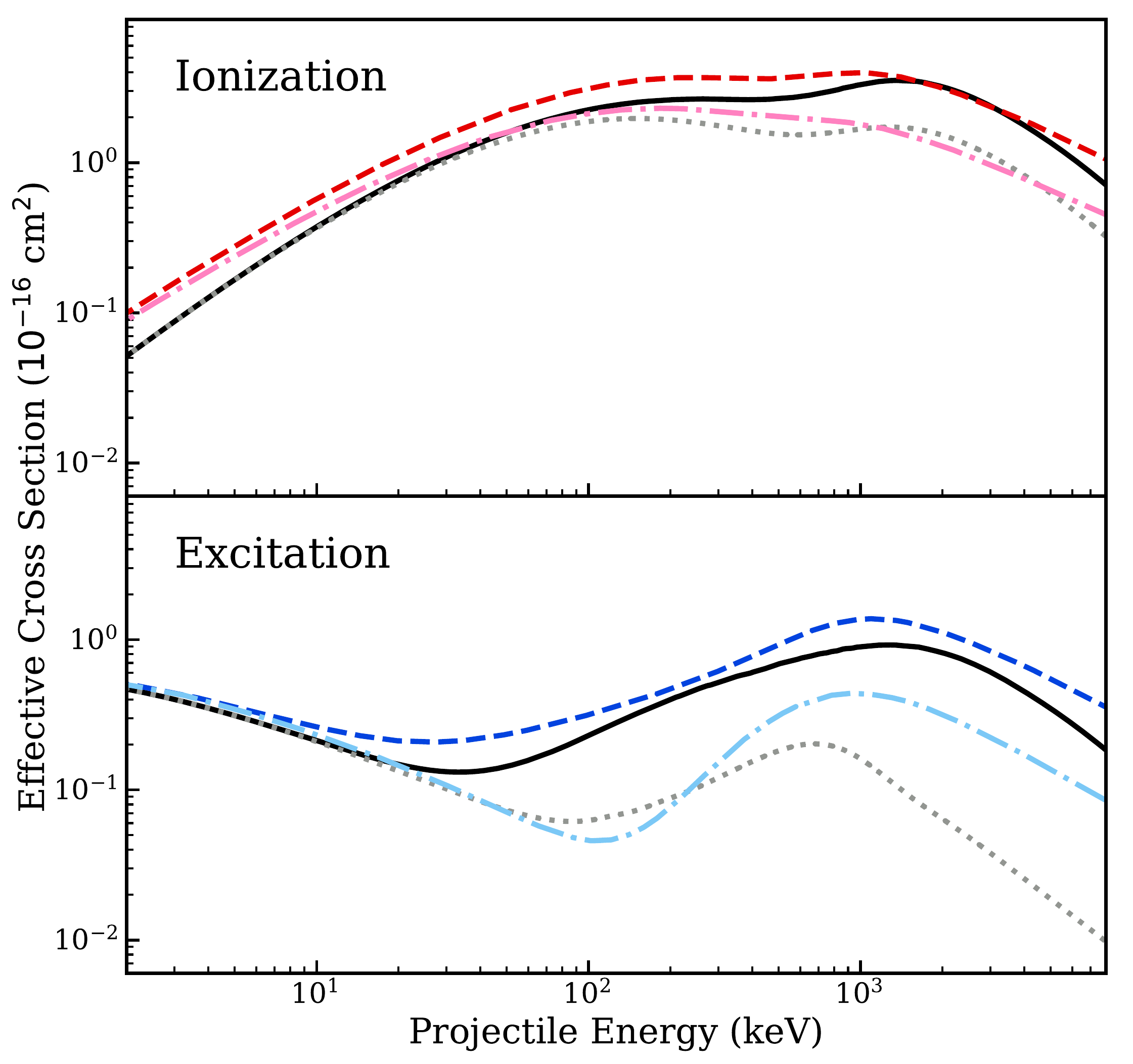}
\caption{\label{fig:model_effective_cross_sections} Effective ionization and excitation cross sections used to compute the energy partitioning of nuclear recoils in superfluid helium. For the model compared to data in this work, we use the curves for ionization with secondary electrons and excitations with secondary electrons (solid black in both panels). We also compare ionizations without secondary electrons and excitations without secondary electrons (dotted gray in both panels) to show the effect of secondary electrons in the model. The cross sections used in this work are compared to Ref.~\cite{Ito13}, from which the curves shown in Fig.~7 are plotted here: ionization without secondary electrons (dash-dotted pink), ionization with secondary electrons (dashed red), excitation without secondary electrons (dash-dotted light blue), excitation with secondary electrons (dashed blue).}
\end{figure}

Atomic helium-helium collision cross section data provide a useful foundation in predicting the scintillation yield of nuclear recoils in superfluid helium \cite{Ito13, Guo:13prd} and the full energy partitioning into various signal channels \cite{Hertel:2019}. In this work, the method used to calculate these predictions follows Ref.~\cite{Hertel:2019} with some modifications and additions. The model described in Ref.~\cite{Hertel:2019} neglected the effect of secondary electrons in producing ionization and excited-state atoms, because this effect was expected to be negligible in the recoil energies of interest below about 100 keV. Figure~\ref{fig:model_effective_cross_sections} shows the effective cross sections for the production of ionization and excited-state atoms without secondary electron contributions in dotted gray in both panels. These are compared to the equivalent cross sections presented in Ref.~\cite{Ito13} (ionization in dash-dotted pink and excitation in dash-dotted light blue). The differences in these curves come from different extrapolations applied to the experimental cross section data.

Since the experimental data in this work were acquired mostly at nuclear recoil energies above 100 keV, the model was updated to consider the effect of secondary electrons on the relative populations of ionization and excited-state atoms in the recoil track. For the secondary electron contribution, we follow a similar procedure as Ref.~\cite{Ito13} by computing the secondary electron energy spectrum with the semiempirical expression from Ref.~\cite{MR1992a}. We find the average secondary electron energy above the ionization energy of 20 eV, and count the average number of ionizations by dividing by the helium electron $W$ value of 43 eV, and we take the ratio of excitations to ionizations to be 0.45 \cite{Ito13}.  The effective ionization and excitation cross sections after the consideration of secondary electron effects are shown in solid black in Fig.~\ref{fig:model_effective_cross_sections}, and again compared to the equivalent curves presented in Ref.~\cite{Ito13} (ionization in dashed red and excitation in dashed blue). We note that the addition of the secondary electron contributions reduces the magnitude of high recoil energy cross section extrapolation differences between this work and Ref.~\cite{Ito13}.

\begin{figure}[ht]
\includegraphics[width=\linewidth]{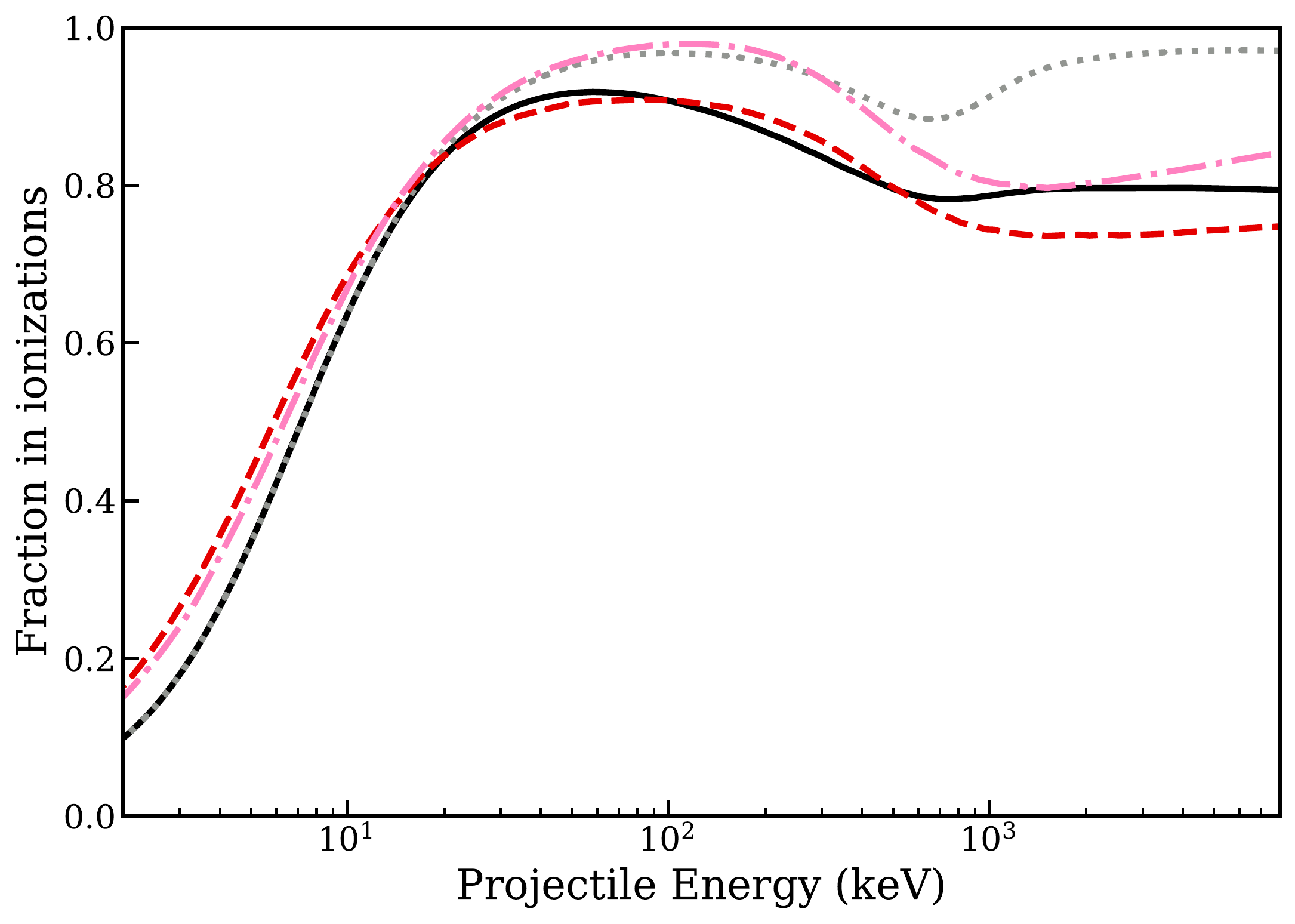}
\caption{\label{fig:model_ionization_fraction} The fraction of Lindhard electronic excitation energy partitioned into ionization, while the remainder is in excitation to singlet or triplet excimers. These curves are computed from the effective ionization and excitation cross sections from Fig.~\ref{fig:model_effective_cross_sections} by dividing the effective ionization cross section by their sum. The curve used for comparison with data in this work (solid black) is computed from the solid black curves in Fig.~\ref{fig:model_effective_cross_sections}, which include the effects of secondary electrons in the effective cross sections. For comparison, we also show curves computed without the effect of secondary electrons (dotted gray), and Ref.~\cite{Ito13} curves without secondary electrons (dash-dotted pink) and with secondary electrons (dashed red).}
\end{figure}

In subsequent steps of the model construction, the effective cross sections for ionization and excitation are summed together and their fractional contribution to the sum gives the ratio of ionization atoms to excited-state atoms as a function of recoil energy. In Fig.~\ref{fig:model_ionization_fraction}, we show that differences in the effective cross sections between our model and the curves from Ref.~\cite{Ito13} do not result in a large modification to this ratio once secondary electrons are taken into account, particularly in the recoil energy range of our experimental data (53.2 to 1090 keV$_\mathrm{nr}$). Finally, this ratio of ionization to excitation is converted to a partitioning of recoil energy into the visible signal channels of singlet excimers, triplet excimers, infrared radiation, and quasiparticle excitations using microphysical assumptions we outlined in Ref.~\cite{Hertel:2019}. Relative to that version of the model, we made some revisions to the characteristic energies for the different types of quanta: singlet and triplet excimers were each assigned average energies 16 eV, the IR channel was assigned 4 eV for each ionization atom and 2 eV for each excited-state atom, and the quasiparticle channel was assigned 8 eV for secondary electron contributions, 2 eV for dimerization, and 4 eV for the dissociation of ground state excimers \cite{GS2019a}. These characteristic energies were applied to the ER partitioning in this work, as well. We also note that the ER model predicts constant fractions for the different signal channels in the energy range probed by our experimental data.

\begin{figure}[ht]
\includegraphics[width=\linewidth]{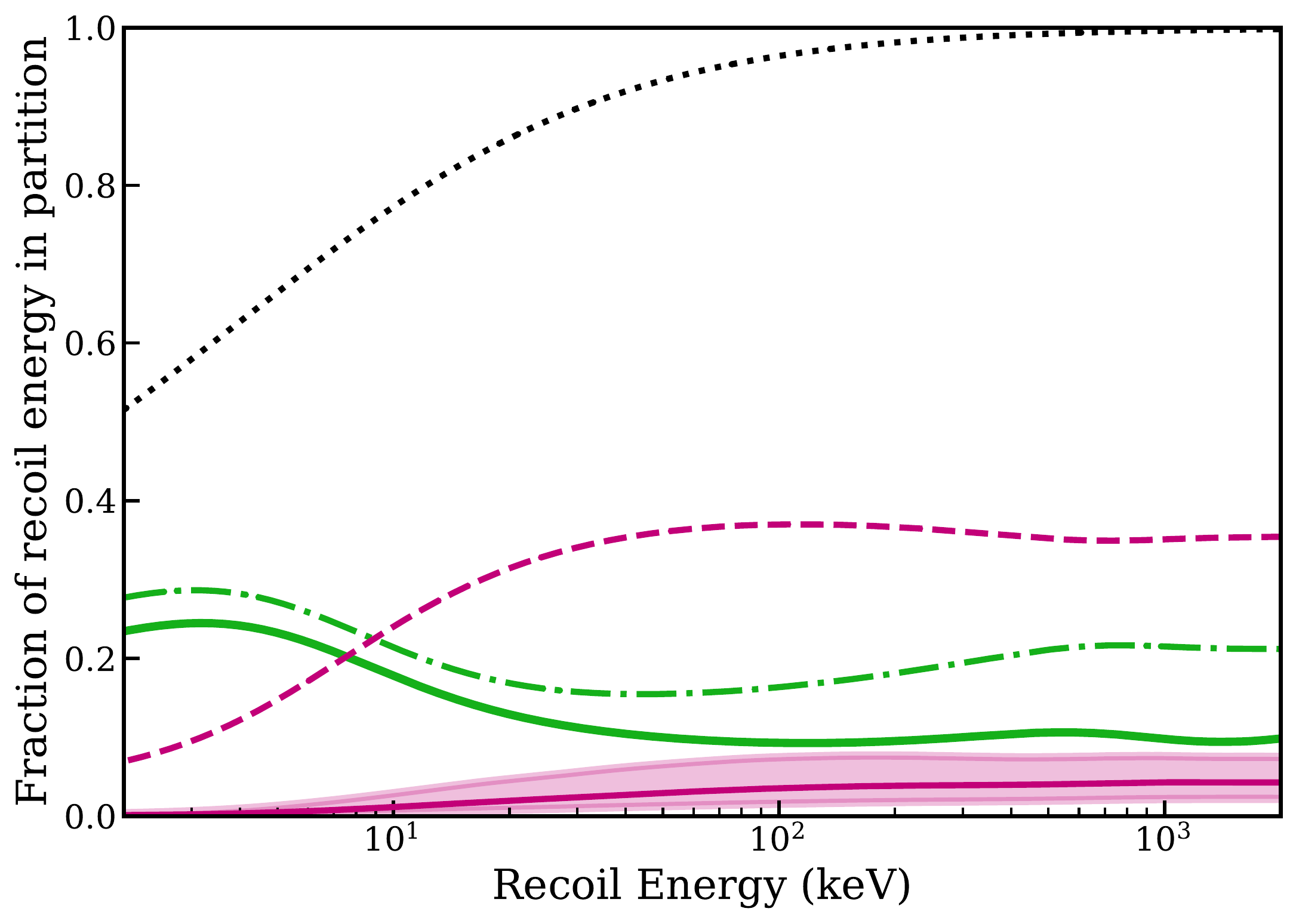}
\caption{\label{fig:model_overview} An overview of the model compared to NR scintillation data. The dotted black line represents the fraction of recoil energy in electronic stopping according to the Lindhard model of the stopping power. The electronic stopping energy is partially partitioned into singlets (dash-dotted green) and triplets (dashed magenta). Due to Penning quenching, only some singlets produce visible scintillation (solid green). Triplets which are ionized by the Penning process and recombine into singlets also produce visible scintillation (solid magenta), with the associated error band in the estimate described in the text.}
\end{figure}

Figure~\ref{fig:model_overview} summarizes the remaining steps in constructing the model that we compare to our experimental data. The recoil energy is partitioned according to the Lindhard fraction of the electronic stopping power in the total stopping power. The Lindhard model predicts that electronic stopping dominates at high recoil energies; for the highest recoil energy datapoint of 1090 keV$_\mathrm{nr}$, electronic stopping represents 99.7\% of the stopping power. Electronic stopping yields both ionization and excited-state atoms along the track in the ratio given by the solid black curve in Fig.~\ref{fig:model_ionization_fraction}. In the absence of an applied electric field, like in this experiment, ions and electrons recombine, yielding excimers in a 1:3 ratio of singlets to triplets. The fraction of the recoil energy found in singlet and triplet excimers after recombination is shown by the dash-dotted green and dashed magenta lines in Fig.~\ref{fig:model_overview}. As discussed in Ref.~\cite{Ito13}, prompt scintillation is quenched by two singlet excimers interacting via the Penning process. Figure~\ref{fig:model_overview} shows the fraction of energy observable as singlet scintillation after the Penning process as the solid green curve, using the same quenching model presented in Ref.~\cite{Ito13}.

Additional scintillation is expected due to Penning processes between two triplet molecules. Recombination of electron-ion pairs formed by this process can produce singlets on a delayed timescale. The rate of scintillation associated with triplet-triplet quenching in a cylindrical track is predicted in Ref.~\cite{King:66}:
\begin{equation}
    \label{eq:king_voltz}
    I'(t) = \frac{k_{f}k_{tt}\tau_{s}}{2\chi_{tt}t_{tt}}\frac{N_T(0)}{[1+\frac{t_d}{2t_{tt}}\mathrm{ln}(1+\frac{t}{t_d})]^2(1+\frac{t}{t_d})}.
\end{equation}
Here the initial number of triplets $N_T(0)$ is predicted by the prompt model and shown as the dashed magenta curve in Fig.~\ref{fig:model_overview}. $\tau_s$ is the singlet lifetime, and the radiative rate parameter $k_f = \tau_s^{-1}$ since each singlet decay produces a photon. The annihilation time is $t_{tt} = \pi r_0^2 L / \chi_{tt} N_T(0)$ where the track radius $r_0$ is taken to be 20~nm, the track length $L$ is the stopping range predicted by ASTAR \cite{NIST_STAR}, and the annihilation constant $\chi_{tt}$ is measured in \cite{Keto:1974} to be $4 \times 10^{-10}$~cm$^3$s$^{-1}$. The diffusion time is $t_{d} = r_0^2 / 4D_T$ where $D_T = 4.2 \times 10^{-4}$~cm$^2$s$^{-1}$ at 2.0~K \cite{Roberts:1973}. Uncertainties on $D_T$ and $\chi_{tt}$ as well as the variation with temperature between 1.6~K and 2.0~K are included as a model uncertainty. More important to the model uncertainty is the parameter $k_{tt} = f\chi_{tt}$ where the fraction $f$ is the number of singlets produced per triplet annihilated. We allow this fraction to vary between $\frac{1}{5}$ in the uncorrelated case where the triplet-to-singlet ratio is 3:1 after recombination, to $\frac{1}{2}$ in the case that a singlet is produced in every recombination, while the median predictions appearing in Figs.~\ref{fig:er_scint_partition} and~\ref{fig:nr_scint_partition} assume the 1:1 ratio observed in geminate recombination \cite{adamsthesis}.

Note that Eq.~(\ref{eq:king_voltz}) only predicts the $t^{-1}$ scintillation component used in fits in the approximation that
\begin{equation*}
    t \gg t_d \quad \mathrm{and} \quad \frac{t_d ln(1+\frac{t}{t_d})}{2t_{tt}} \ll 1.
\end{equation*} 
The former approximation is a good one for the microsecond timescales measured here; the prediction for $t_d$ is about 3~ns. However, the latter approximation is less accurate. For the NR energies studied here, the above fraction is expected to approach or slightly exceed 1 at the end of the event window, with the worst violation occurring for NR recoils around 1~MeV. Fits with the full functional form in Eq.~(\ref{eq:king_voltz}) with $t_d$ and $t_{tt}$ fixed to their predicted values were performed for comparison and did not significantly affect goodness-of-fit or the scintillation partitioning results presented here. Improved statistical uncertainty, better resolution between the prompt pulse and delayed scintillation, and a longer event window could allow measurements of $t_d$ and $t_{tt}$ through fits to the full time dependence in Eq.~(\ref{eq:king_voltz}).

Modeling in Ref.~\cite{Hertel:2019} also considered the partitioning of recoil energy into infrared radiation. While the infrared channel is still present in the model described in this paper, it lies outside of the sensitivity range of the PMTs used in this experiment, so we do not consider it in our comparison to data. 

\subsection{Relative light yield}

\begin{figure}[ht]
\includegraphics[width=\linewidth]{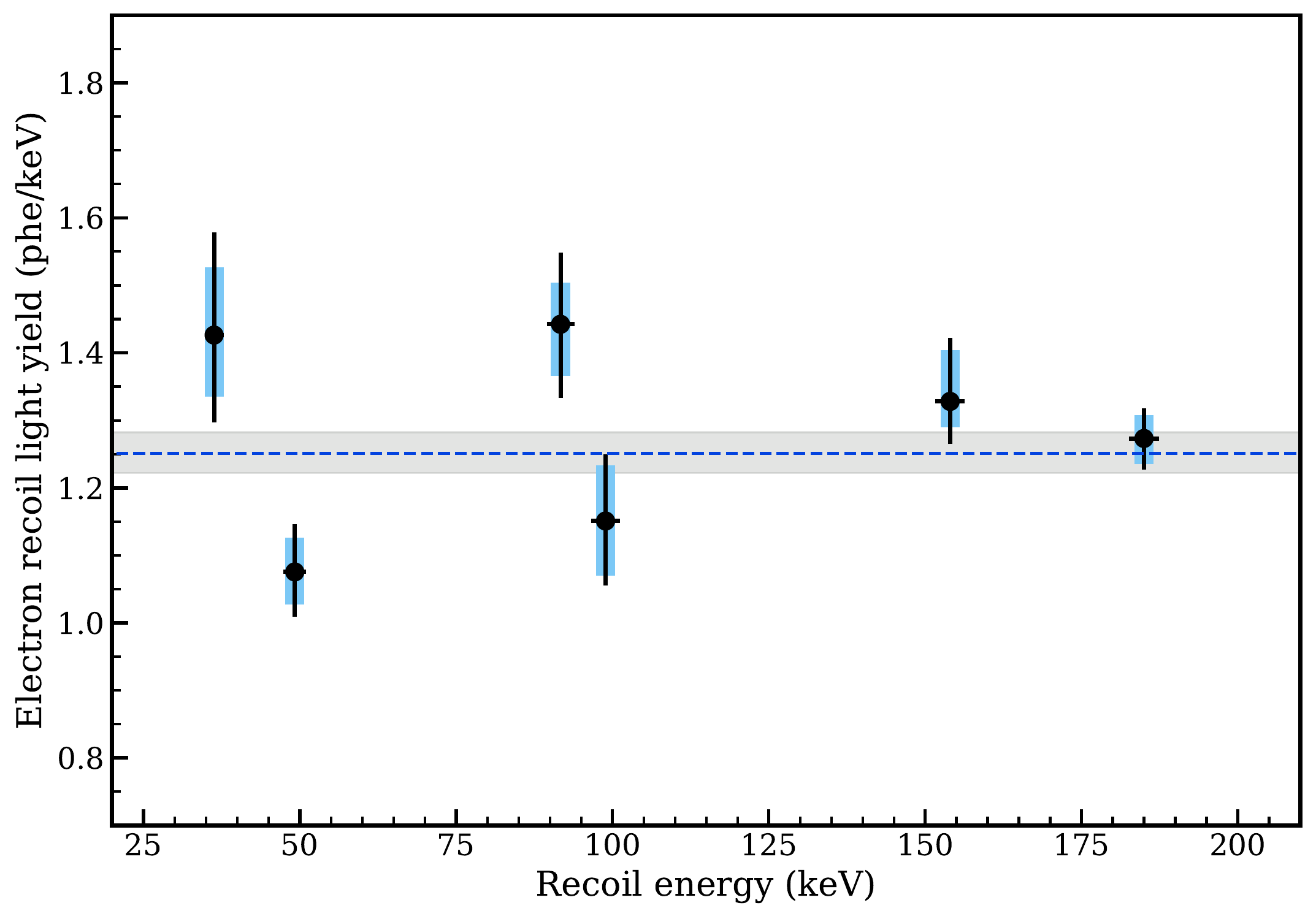}
\caption{\label{fig:ER_LY_plot} The measured ER signal sizes for the datasets shown in Fig.~\ref{fig:ER_fits} (points) and weighted mean ER signal size (blue dashed line). The vertical extent of the blue bands represents the systematic uncertainty associated with the measurement, and the black line is the total error from summing the systematic and statistical errors in quadrature. Table~\ref{tab:errors} lists these errors for each dataset. Horizontal black lines are the error on the recoil energy associated with the dataset also listed in Table~\ref{tab:errors}; these are hidden by the markers for several points. The gray band represents the error on the weighted mean ER signal size.}
\end{figure}

\begin{figure}[ht]
\includegraphics[width=\linewidth]{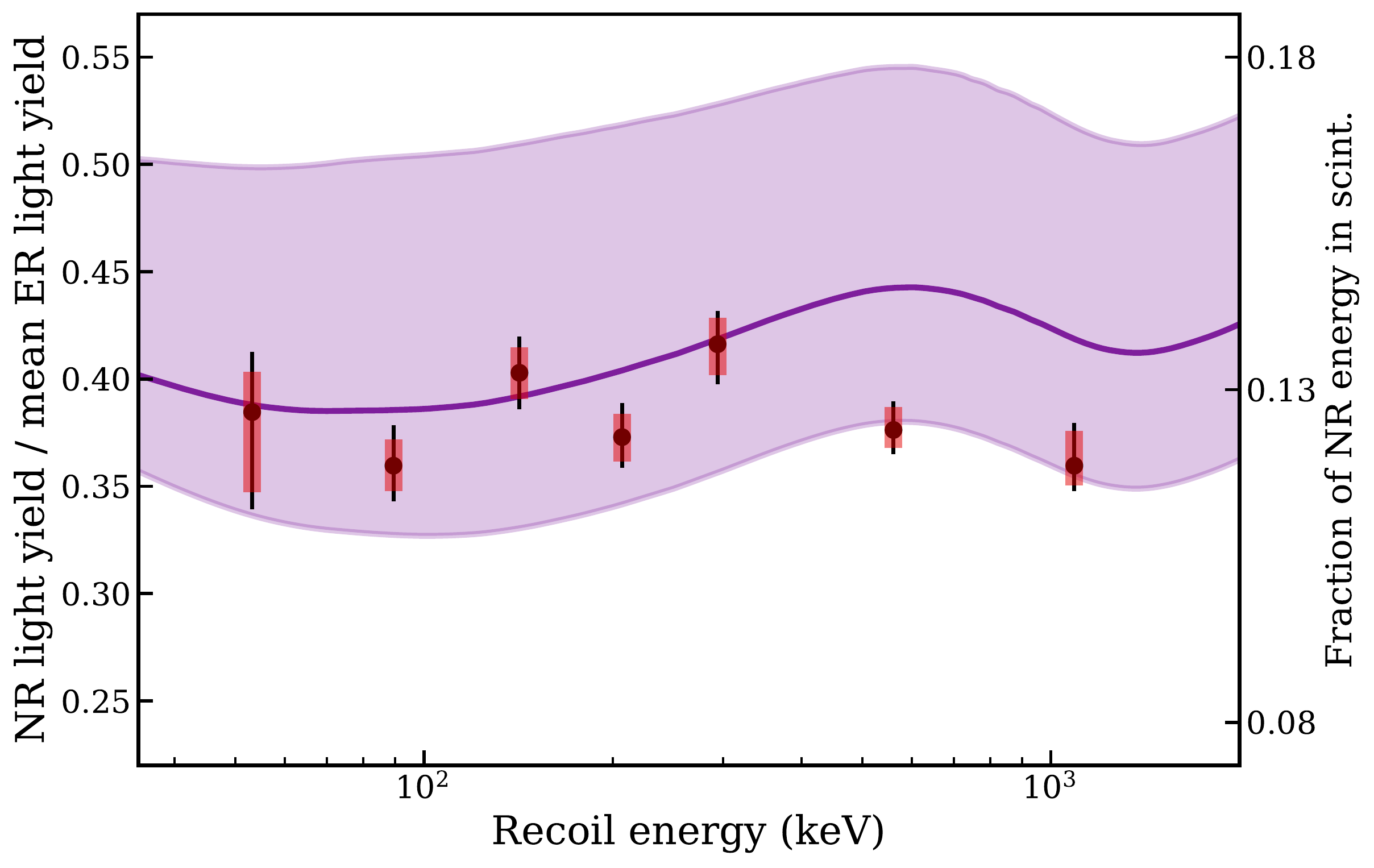}
\caption{\label{fig:relative_light_yield} The relative light yields measured in this experiment, computed as the NR signal size parameters in Table~\ref{tab:errors} divided by the average ER signal size shown in Fig.~\ref{fig:ER_LY_plot}. As in Fig.~\ref{fig:ER_LY_plot}, the red bars represent systematic uncertainties and the thin black lines the total error associated with the measurement at each energy. On this scale, all horizontal error bars are hidden by the markers; they are listed in Table~\ref{tab:errors}. The experimental results are compared to the predicted behavior from the semiempirical model (purple) described in Sec.~\ref{ssec:model_description}, computed here as the sum of the singlet and triplet contribution to the scintillation signal. The right y-axis on this plot represents the fraction of NR energy recovered as scintillation assuming the ER light yield predicted by the model.}
\end{figure}

The ER signal size values from Table~\ref{tab:errors} are plotted in Fig.~\ref{fig:ER_LY_plot} along with their mean value, $1.25^{+0.03}_{-0.03}$ phe/keV. The NR signal size values were divided by the mean ER signal size and plotted in Fig.~\ref{fig:relative_light_yield}. Since the data were taken with the same target volume, this division should account for any geometric effects on the light yield and facilitate comparison to more fundamental physical models. The model prediction for the relative light yield is also plotted in Fig.~\ref{fig:relative_light_yield} as the solid purple line; the ER fraction of recoil energy in scintillation is flat across the energy range of the measurements and predicted to be 0.32. This model prediction curve is the sum of the prompt scintillation curve from singlet decays (the solid green line in Fig.~\ref{fig:model_overview}) and the delayed scintillation model for our event window length (the solid magenta line in Fig.~\ref{fig:model_overview}). The level of agreement between the data and model prediction is quite encouraging, given that no fitting was performed in their comparison.

An interesting feature in the model curve is a rise in the light yield below our lowest measured energy of 53 keV$_\mathrm{nr}$. This prediction provides ample motivation for nuclear recoil light yield measurements at lower energies, to further constrain the microphysical modeling of particle interactions in helium. We also note that these measurements were taken at a single temperature of 1.75~K, while future dark matter detection schemes anticipate operating temperatures in the tens of mK \cite{Hertel:2019}. The expected temperature dependence of various microphysical processes further motivates future measurements like the ones presented in this paper at lower temperatures.

\subsection{Delayed components}

\begin{figure}[ht]
\includegraphics[width=\linewidth]{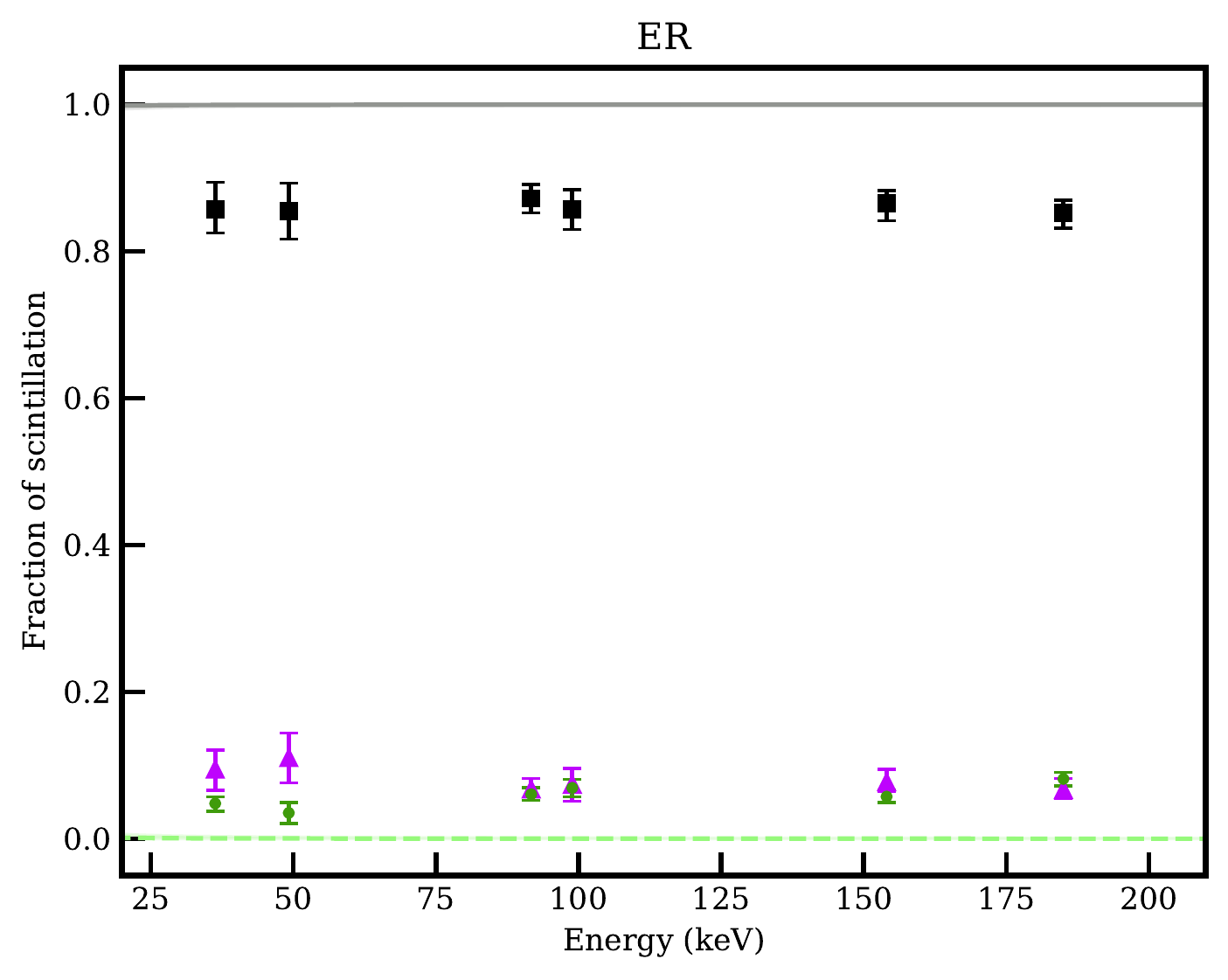}
\caption{\label{fig:er_scint_partition} Fraction of ER scintillation light in the prompt ($<$640~ns, black square), exponential (purple triangle), and 1/t (green circle) components. Lines are the fraction of total scintillation predicted from singlets (solid gray) and triplets (dashed green); the error bands associated with the predictions are smaller than the width of the plotted lines.}
\end{figure}

\begin{figure}[ht]
\includegraphics[width=\linewidth]{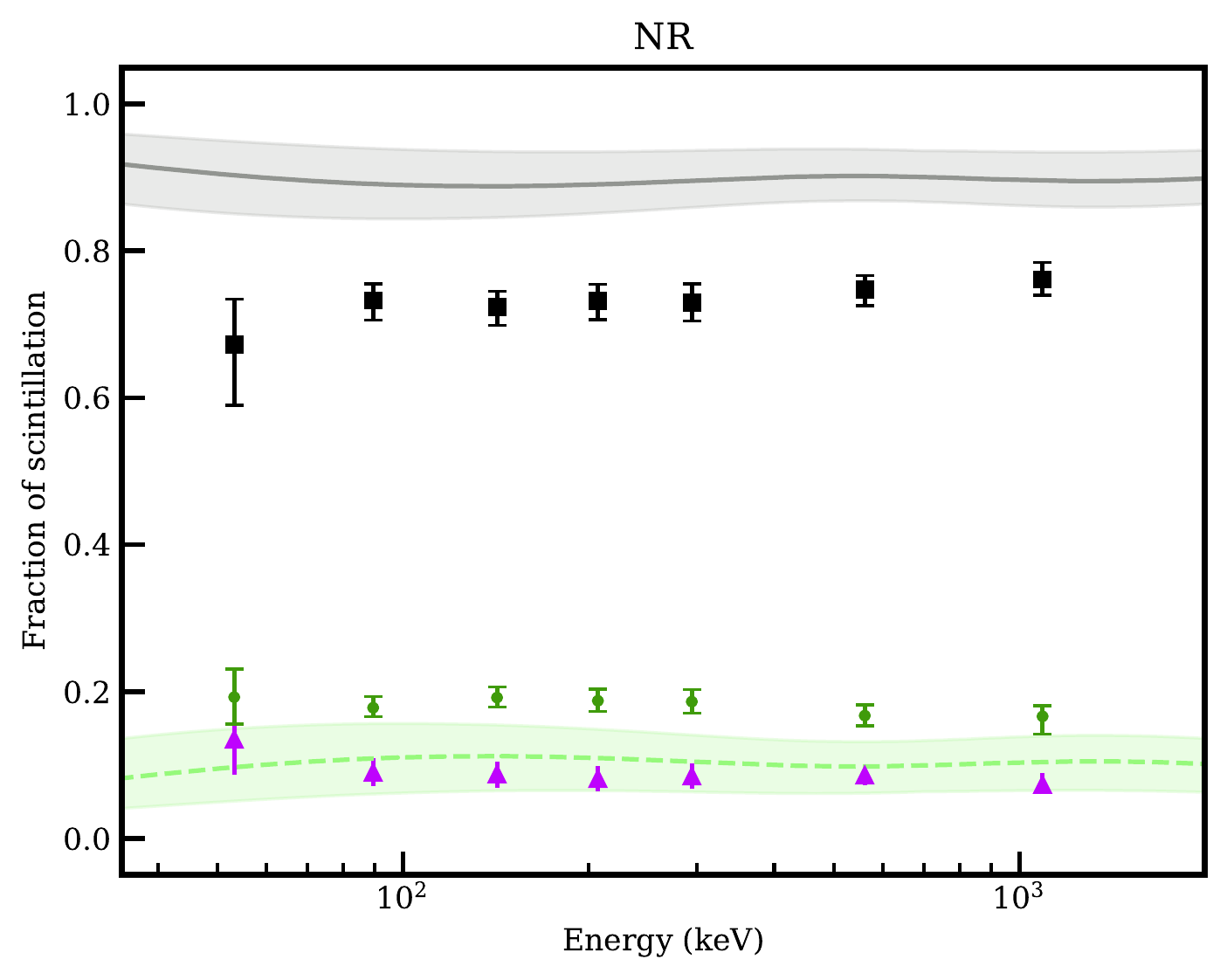}
\caption{\label{fig:nr_scint_partition} Fraction of NR scintillation light in the prompt ($<$640~ns, black square), exponential (purple triangle), and 1/t (green circle) components. Bands with central lines are the fraction of total scintillation from singlets (gray, solid central line) and triplets (green, dashed central line), by taking the fraction of each in the sum of the corresponding solid curves in Fig.~\ref{fig:model_overview}.}
\end{figure}

The partitioning results from the delayed scintillation analysis are shown in Figs.~\ref{fig:er_scint_partition}~and~\ref{fig:nr_scint_partition} for ER data and NR data respectively. Comparing these results, it is clear that the prompt fraction was found to be higher for ERs compared to NRs. This result was due to a smaller $t^{-1}$ component for ERs, whereas the contribution of the exponential component was similar for both recoil types. Since the excitations from NRs are confined on smaller scales spatially than from ERs, the bimolecular processes producing the $t^{-1}$ component are more prominent in the NR case. The difference in scintillation timing between ERs and NRs could be useful for ER/NR discrimination. We do not quantify the discrimination power here, and in a detector that is sensitive to additional channels such as phonons/rotons as well as, discrimination by scintillation timing may be secondary to discrimination by partitioning among the signal channels.

Figures~\ref{fig:er_scint_partition} and~\ref{fig:nr_scint_partition} also show model curves to compare to the data. These are simply the fraction that the singlet and triplet scintillation curves contribute to the total expected scintillation. The gray curve is the singlet scintillation fraction, comparable to the prompt component in the experimental data, while the green curve comes from triplets scintillating by way of the Penning process, the expected source of the $t^{-1}$ component also in green. The exponential component does not have corresponding partition in the model since its source is unknown. However, the sum of the exponential and prompt data points would provide fair agreement with the singlet curve. This comparison stands as some preliminary evidence for the hypothesis presented in Ref.~\cite{McKinsey:03pra} that the exponential component is due to reactions between metastable He($2^{1}S$) and ground state helium atoms. 

The partitioning of scintillation among the three components identified shows no significant energy dependence for either NRs or ERs, nor does the model predict strong energy dependence. Below the energies used here, down to 10~keV, the $t^{-1}$ is expected to decrease substantially for NR recoils and increase for ER recoils. As is the case with the total scintillation comparison, more data at lower recoil energies would be valuable to compare to these predictions. 

The lifetime of the exponential component does, on the other hand, appear to vary with recoil energy for NRs in the experimental data. While the lifetime is $\approx 1.4~\mu s$ for ERs at all energies and for lower energy NRs, similar to the $\approx 1.6~\mu s$ previously reported, it increases with energy to $\approx 2.5~\mu s$ for the two highest NR energies. One hypothesis for this behavior is that the lifetime at small excimer density is constant at the 1.4~$\mu$s value observed for ERs, but that additional time dependence is introduced by quenching of the signal. As the density of excitations in the helium decreases over time, the signal may become less quenched, stretching the lifetime when fit with an exponential. Assuming the excited state responsible for this component is quenched by the bath of triplets with the same annihilation constant $\chi_{tt} = 4 \times 10^{-10}$~cm$^3$s$^{-1}$ measured for triplet-triplet interactions, the signal will initially be quenched since $\chi_{tt}n_T(0) > 1/\tau_{exp}$ with an initial triplet concentration above $10^{17} / \mathrm{cm}^3$ for NR energies above 20~keV, taking the lifetime to be $\tau_{exp} = 1.4\ \mu s$. The triplet concentration decreases primarily due to diffusion; with $D_T \approx 7 \times 10^{-4}$ cm$^2/$s and an initial cylindrical track with radius $r_0$ of 20~nm, we can estimate the timescale $t_q$ at which the signal transitions to unquenched: 
\begin{equation}
    t_q = \frac{\chi_{tt}\tau_{exp}n_T(0)r_0^2}{D_T}.
\end{equation}
At the lowest NR energy used here, the initial triplet concentration in our model is $\approx 1.7 \times 10^{17} / \mathrm{cm}^3$, so $t_q \approx 600$~ns; there is little quenching in the fit window, which begins at around 640~ns. In this case, we would expect to reconstruct the lifetime at low excimer density well. However, at the highest energy, $n_T(0)\approx 5.2 \times 10^{17} / \mathrm{cm}^3$ so $t_q \approx 1.8~\mu$s. Quenching extends well into the fit window, but disappears before the exponential has decayed away entirely, stretching the fit lifetime. The initial concentration increases monotonically with energy in this range, so it appears plausible that the trend in fit lifetime is due to this effect. Moreover, there is little tension with the lower lifetimes found in the literature using higher energy alpha sources, since the initial concentration decreases again above 1-2~MeV. Time dependence due to quenching could be investigated further by varying temperature, which strongly affects the diffusion constant $D_T$ \cite{Roberts:1973}.

It should be noted that delayed photon emission from TPB may in principle contribute to the observed scintillation time dependence. Reference~\cite{Segreto_2015} has measured TPB emission on similar timescales, and even attributes the delayed TPB emission to the same triplet-triplet quenching mechanism discussed here that predicts the $t^{-1}$ component, but between photoionized TPB molecules rather than triplets in the helium. In Ref~\cite{McKinsey:03pra}, however, it was observed that both delayed components differed between cold helium gas versus liquid helium, demonstrating that the helium itself was playing a role in the delayed scintillation. The differing strength of the $t^{-1}$ component between ERs and NRs also is not well explained by delayed emission from the TPB. A partial contribution to the delayed components could boost the $t^{-1}$ fraction above the predictions, particularly in the ER case where Eq.~(\ref{eq:king_voltz}) predicts only subpercent contributions from triplet-triplet quenching due to low excimer density.

\section{Conclusion}
\label{sec:conclusion}
We have used fixed-angle gamma and neutron scattering to measure the relative light yield of superfluid $^4$He with PMTs immersed in the 1.75~K superfluid for a range of recoil energies. We observed a signal size of $1.25^{+0.03}_{-0.03}$ phe/keV for Compton scatter recoils in the 10s and 100s of keV$_\mathrm{ee}$. Additionally, we presented the first measurements of the superfluid $^4$He relative light yield down to 53 keV$_\mathrm{nr}$. Our measurements show encouraging agreement with a semiempirical model predicting the ratio of NR and ER light yields without any direct fitting to our data. The predicted increase in the relative light yield of NRs with decreasing energies is certainly interesting for future experimentation.

We additionally used our experimental data to explore the time dependence of delayed scintillation components. The analyzed data demonstrated exponential and $t^{-1}$ components consistent with previous measurements. We also discussed a model for the $t^{-1}$ yield. The fraction of scintillation in that component appeared fairly consistent with the highest prediction in the NR case, and higher than predicted in the ER case. The previously studied exponential component was newly found to have some energy dependence in the best fit lifetime value. Observation of the delayed scintillation in superfluid helium not only offers an ER/NR discrimination tool, but also a window into the helium microphysics; the multiple components of time dependence shed light on the excited species and their interactions.

Measurements such as the ones presented here serve as valuable input to microphysical models of particle interactions in superfluid helium. Furthermore, they serve as further evidence for the promising qualities of superfluid helium as a particle detector target material.

\begin{acknowledgments}
This work was supported in part by DOE Grant No. DE-SC0019319, and DOE Quantum Information Science Enabled Discovery (QuantISED) for High Energy Physics (KA2401032). W.G. acknowledges the support from the National High Magnetic Field Laboratory at Florida State University, which is funded through the NSF Cooperative Agreement No. DMR-1644779 and the state of Florida. V.V. is supported by a DOE Graduate Instrumentation Research Award.  This material is based upon work supported by the National Science Foundation Graduate Research Fellowship under Grant No. DGE 1106400. This research used the Savio computational cluster resource provided by the Berkeley Research Computing program at the University of California, Berkeley (supported by the UC Berkeley Chancellor, Vice Chancellor for Research, and Chief Information Officer). We thank Hans Kraus from University of Oxford for useful discussion.
\end{acknowledgments}

\bibliography{Helium_Bib_12_2020.bib}

\begin{thebibliography}{42}%
\makeatletter
\providecommand \@ifxundefined [1]{%
 \@ifx{#1\undefined}
}%
\providecommand \@ifnum [1]{%
 \ifnum #1\expandafter \@firstoftwo
 \else \expandafter \@secondoftwo
 \fi
}%
\providecommand \@ifx [1]{%
 \ifx #1\expandafter \@firstoftwo
 \else \expandafter \@secondoftwo
 \fi
}%
\providecommand \natexlab [1]{#1}%
\providecommand \enquote  [1]{``#1''}%
\providecommand \bibnamefont  [1]{#1}%
\providecommand \bibfnamefont [1]{#1}%
\providecommand \citenamefont [1]{#1}%
\providecommand \href@noop [0]{\@secondoftwo}%
\providecommand \href [0]{\begingroup \@sanitize@url \@href}%
\providecommand \@href[1]{\@@startlink{#1}\@@href}%
\providecommand \@@href[1]{\endgroup#1\@@endlink}%
\providecommand \@sanitize@url [0]{\catcode `\\12\catcode `\$12\catcode
  `\&12\catcode `\#12\catcode `\^12\catcode `\_12\catcode `\%12\relax}%
\providecommand \@@startlink[1]{}%
\providecommand \@@endlink[0]{}%
\providecommand \url  [0]{\begingroup\@sanitize@url \@url }%
\providecommand \@url [1]{\endgroup\@href {#1}{\urlprefix }}%
\providecommand \urlprefix  [0]{URL }%
\providecommand \Eprint [0]{\href }%
\providecommand \doibase [0]{http://dx.doi.org/}%
\providecommand \selectlanguage [0]{\@gobble}%
\providecommand \bibinfo  [0]{\@secondoftwo}%
\providecommand \bibfield  [0]{\@secondoftwo}%
\providecommand \translation [1]{[#1]}%
\providecommand \BibitemOpen [0]{}%
\providecommand \bibitemStop [0]{}%
\providecommand \bibitemNoStop [0]{.\EOS\space}%
\providecommand \EOS [0]{\spacefactor3000\relax}%
\providecommand \BibitemShut  [1]{\csname bibitem#1\endcsname}%
\let\auto@bib@innerbib\@empty
\bibitem [{\citenamefont {Adams}\ \emph {et~al.}(1996)\citenamefont {Adams}
  \emph {et~al.}}]{Adams:1996}%
  \BibitemOpen
  \bibfield  {author} {\bibinfo {author} {\bibfnamefont {J.~S.}\ \bibnamefont
  {Adams}} \emph {et~al.},\ }in\ \href@noop {} {\emph {\bibinfo {booktitle}
  {Dark matter in cosmology, quantum measurements, experimental
  gravitation}}},\ \bibinfo {editor} {edited by\ \bibinfo {editor}
  {\bibfnamefont {R.}~\bibnamefont {Ansari}}, \bibinfo {editor} {\bibfnamefont
  {J.}~\bibnamefont {Tran Thanh~Van}}, \ and\ \bibinfo {editor} {\bibfnamefont
  {Y.}~\bibnamefont {Giraud-Heraud}}},\ \bibinfo {organization} {Ed.
  Frontieres}\ (\bibinfo  {publisher} {Ed. Frontieres},\ \bibinfo {address}
  {Gif-Sur-Yvette},\ \bibinfo {year} {1996})\ pp.\ \bibinfo {pages}
  {131--136}\BibitemShut {NoStop}%
\bibitem [{\citenamefont {Guo}\ and\ \citenamefont
  {McKinsey}(2013)}]{Guo:13prd}%
  \BibitemOpen
  \bibfield  {author} {\bibinfo {author} {\bibfnamefont {W.}~\bibnamefont
  {Guo}}\ and\ \bibinfo {author} {\bibfnamefont {D.~N.}\ \bibnamefont
  {McKinsey}},\ }\href {\doibase 10.1103/PhysRevD.87.115001} {\bibfield
  {journal} {\bibinfo  {journal} {Phys. Rev. D}\ }\textbf {\bibinfo {volume}
  {87}},\ \bibinfo {pages} {115001} (\bibinfo {year} {2013})}\BibitemShut
  {NoStop}%
\bibitem [{\citenamefont {Ito}\ and\ \citenamefont {Seidel}(2013)}]{Ito13}%
  \BibitemOpen
  \bibfield  {author} {\bibinfo {author} {\bibfnamefont {T.}~\bibnamefont
  {Ito}}\ and\ \bibinfo {author} {\bibfnamefont {G.}~\bibnamefont {Seidel}},\
  }\href {\doibase 10.1103/PhysRevC.88.025805} {\bibfield  {journal} {\bibinfo
  {journal} {Phys. Rev. C}\ }\textbf {\bibinfo {volume} {88}},\ \bibinfo
  {pages} {025805} (\bibinfo {year} {2013})}\BibitemShut {NoStop}%
\bibitem [{\citenamefont {Maris}\ \emph {et~al.}(2017)\citenamefont {Maris},
  \citenamefont {Seidel},\ and\ \citenamefont {Stein}}]{Mar:2017}%
  \BibitemOpen
  \bibfield  {author} {\bibinfo {author} {\bibfnamefont {H.}~\bibnamefont
  {Maris}}, \bibinfo {author} {\bibfnamefont {G.}~\bibnamefont {Seidel}}, \
  and\ \bibinfo {author} {\bibfnamefont {D.}~\bibnamefont {Stein}},\ }\href
  {https://doi.org/10.1103/PhysRevLett.119.181303} {\bibfield  {journal}
  {\bibinfo  {journal} {Phys. Rev. Lett.}\ }\textbf {\bibinfo {volume} {119}},\
  \bibinfo {pages} {181303} (\bibinfo {year} {2017})}\BibitemShut {NoStop}%
\bibitem [{\citenamefont {Hertel}\ \emph {et~al.}(2019)\citenamefont {Hertel},
  \citenamefont {Biekert}, \citenamefont {Lin}, \citenamefont {Velan},\ and\
  \citenamefont {McKinsey}}]{Hertel:2019}%
  \BibitemOpen
  \bibfield  {author} {\bibinfo {author} {\bibfnamefont {S.~A.}\ \bibnamefont
  {Hertel}}, \bibinfo {author} {\bibfnamefont {A.}~\bibnamefont {Biekert}},
  \bibinfo {author} {\bibfnamefont {J.}~\bibnamefont {Lin}}, \bibinfo {author}
  {\bibfnamefont {V.}~\bibnamefont {Velan}}, \ and\ \bibinfo {author}
  {\bibfnamefont {D.~N.}\ \bibnamefont {McKinsey}},\ }\href {\doibase
  http://dx.doi.org/10.1103/PhysRevD.100.092007} {\bibfield  {journal}
  {\bibinfo  {journal} {Phys. Rev. D}\ }\textbf {\bibinfo {volume} {100}},\
  \bibinfo {pages} {092007} (\bibinfo {year} {2019})}\BibitemShut {NoStop}%
\bibitem [{\citenamefont {Liao}\ \emph {et~al.}(2021)\citenamefont {Liao},
  \citenamefont {Gao}, \citenamefont {Liang}, \citenamefont {Peng},
  \citenamefont {Zhang},\ and\ \citenamefont {Zhang}}]{JL2021a}%
  \BibitemOpen
  \bibfield  {author} {\bibinfo {author} {\bibfnamefont {J.}~\bibnamefont
  {Liao}}, \bibinfo {author} {\bibfnamefont {Y.}~\bibnamefont {Gao}}, \bibinfo
  {author} {\bibfnamefont {Z.}~\bibnamefont {Liang}}, \bibinfo {author}
  {\bibfnamefont {Z.}~\bibnamefont {Peng}}, \bibinfo {author} {\bibfnamefont
  {L.}~\bibnamefont {Zhang}}, \ and\ \bibinfo {author} {\bibfnamefont
  {L.}~\bibnamefont {Zhang}},\ }\href@noop {} {\  (\bibinfo {year} {2021})},\
  \Eprint {http://arxiv.org/abs/2103.02161} {arXiv:2103.02161} \BibitemShut
  {NoStop}%
\bibitem [{\citenamefont {Alexander}\ \emph {et~al.}(2016)\citenamefont
  {Alexander} \emph {et~al.}}]{Alexander:2016aln}%
  \BibitemOpen
  \bibfield  {author} {\bibinfo {author} {\bibfnamefont {J.}~\bibnamefont
  {Alexander}} \emph {et~al.}\ }(\bibinfo {year} {2016})\ \Eprint
  {http://arxiv.org/abs/1608.08632} {arXiv:1608.08632} \BibitemShut {NoStop}%
\bibitem [{\citenamefont {Battaglieri}\ \emph {et~al.}(2017)\citenamefont
  {Battaglieri} \emph {et~al.}}]{Bat:2017}%
  \BibitemOpen
  \bibfield  {author} {\bibinfo {author} {\bibnamefont {Battaglieri}} \emph
  {et~al.},\ }\href@noop {} {\  (\bibinfo {year} {2017})},\ \Eprint
  {http://arxiv.org/abs/1707.04591} {arXiv:1707.04591} \BibitemShut {NoStop}%
\bibitem [{brn(2018)}]{brn_report}%
  \BibitemOpen
  \href@noop {} {\enquote {\bibinfo {title} {{DOE} report: Basic research needs
  workshop for dark matter small projects new initiatives},}\ }\bibinfo
  {howpublished}
  {\url{https://science.osti.gov/hep/Community-Resources/Reports}} (\bibinfo
  {year} {2018})\BibitemShut {NoStop}%
\bibitem [{\citenamefont {McKinsey}\ \emph {et~al.}(2003)\citenamefont
  {McKinsey} \emph {et~al.}}]{McKinsey:03pra}%
  \BibitemOpen
  \bibfield  {author} {\bibinfo {author} {\bibfnamefont {D.~N.}\ \bibnamefont
  {McKinsey}} \emph {et~al.},\ }\href {\doibase
  https://doi.org/10.1103/PhysRevA.67.062716} {\bibfield  {journal} {\bibinfo
  {journal} {Phys. Rev. A}\ }\textbf {\bibinfo {volume} {67}},\ \bibinfo
  {pages} {062716} (\bibinfo {year} {2003})}\BibitemShut {NoStop}%
\bibitem [{\citenamefont {Huffman}\ \emph {et~al.}(2000)\citenamefont {Huffman}
  \emph {et~al.}}]{Huffman:00n}%
  \BibitemOpen
  \bibfield  {author} {\bibinfo {author} {\bibfnamefont {P.~R.}\ \bibnamefont
  {Huffman}} \emph {et~al.},\ }\href {\doibase 10.1038/47444} {\bibfield
  {journal} {\bibinfo  {journal} {Nature}\ }\textbf {\bibinfo {volume} {403}},\
  \bibinfo {pages} {62} (\bibinfo {year} {2000})}\BibitemShut {NoStop}%
\bibitem [{\citenamefont {Lanou}\ \emph {et~al.}(1987)\citenamefont {Lanou},
  \citenamefont {Maris},\ and\ \citenamefont {Seidel}}]{Lanou:87prl}%
  \BibitemOpen
  \bibfield  {author} {\bibinfo {author} {\bibfnamefont {R.~E.}\ \bibnamefont
  {Lanou}}, \bibinfo {author} {\bibfnamefont {H.~J.}\ \bibnamefont {Maris}}, \
  and\ \bibinfo {author} {\bibfnamefont {G.~M.}\ \bibnamefont {Seidel}},\
  }\href {https://doi.org/10.1103/PhysRevLett.58.2498} {\bibfield  {journal}
  {\bibinfo  {journal} {Phys. Rev. Lett.}\ }\textbf {\bibinfo {volume} {58}},\
  \bibinfo {pages} {2498} (\bibinfo {year} {1987})}\BibitemShut {NoStop}%
\bibitem [{\citenamefont {McKinsey}\ and\ \citenamefont
  {Doyle}(2000)}]{McKinsey:2000}%
  \BibitemOpen
  \bibfield  {author} {\bibinfo {author} {\bibfnamefont {D.}~\bibnamefont
  {McKinsey}}\ and\ \bibinfo {author} {\bibfnamefont {J.}~\bibnamefont
  {Doyle}},\ }\href {\doibase https://doi.org/10.1023/A:1004690906370}
  {\bibfield  {journal} {\bibinfo  {journal} {J. Low Temp. Phys.}\ }\textbf
  {\bibinfo {volume} {118}},\ \bibinfo {pages} {153} (\bibinfo {year}
  {2000})}\BibitemShut {NoStop}%
\bibitem [{\citenamefont {Golub}\ and\ \citenamefont
  {Lamoreaux}(1994)}]{Golub:94pr}%
  \BibitemOpen
  \bibfield  {author} {\bibinfo {author} {\bibfnamefont {R.}~\bibnamefont
  {Golub}}\ and\ \bibinfo {author} {\bibfnamefont {S.~K.}\ \bibnamefont
  {Lamoreaux}},\ }\href {https://doi.org/10.1016/0370-1573(94)90084-1}
  {\bibfield  {journal} {\bibinfo  {journal} {Phys. Rep.}\ }\textbf {\bibinfo
  {volume} {237}},\ \bibinfo {pages} {1} (\bibinfo {year} {1994})}\BibitemShut
  {NoStop}%
\bibitem [{\citenamefont {Ahmed}\ \emph {et~al.}(2019)\citenamefont {Ahmed}
  \emph {et~al.}}]{Ahmed:2019}%
  \BibitemOpen
  \bibfield  {author} {\bibinfo {author} {\bibfnamefont {M.}~\bibnamefont
  {Ahmed}} \emph {et~al.},\ }\href {\doibase 10.1088/1748-0221/14/11/p11017}
  {\bibfield  {journal} {\bibinfo  {journal} {Journal of Instrumentation}\
  }\textbf {\bibinfo {volume} {14}},\ \bibinfo {pages} {P11017} (\bibinfo
  {year} {2019})}\BibitemShut {NoStop}%
\bibitem [{\citenamefont {Scholberg}(2006)}]{Scholberg:06prd}%
  \BibitemOpen
  \bibfield  {author} {\bibinfo {author} {\bibfnamefont {K.}~\bibnamefont
  {Scholberg}},\ }\href {\doibase https://doi.org/10.1103/PhysRevD.73.033005}
  {\bibfield  {journal} {\bibinfo  {journal} {Physical Review D}\ }\textbf
  {\bibinfo {volume} {73}},\ \bibinfo {pages} {033005} (\bibinfo {year}
  {2006})}\BibitemShut {NoStop}%
\bibitem [{\citenamefont {Cadeddu}\ \emph {et~al.}(2019)\citenamefont
  {Cadeddu}, \citenamefont {Dordei}, \citenamefont {Giunti}, \citenamefont
  {Kouzakov}, \citenamefont {Picciau},\ and\ \citenamefont
  {Studenikin}}]{Cadeddu:2019}%
  \BibitemOpen
  \bibfield  {author} {\bibinfo {author} {\bibfnamefont {M.}~\bibnamefont
  {Cadeddu}}, \bibinfo {author} {\bibfnamefont {F.}~\bibnamefont {Dordei}},
  \bibinfo {author} {\bibfnamefont {C.}~\bibnamefont {Giunti}}, \bibinfo
  {author} {\bibfnamefont {K.~A.}\ \bibnamefont {Kouzakov}}, \bibinfo {author}
  {\bibfnamefont {E.}~\bibnamefont {Picciau}}, \ and\ \bibinfo {author}
  {\bibfnamefont {A.~I.}\ \bibnamefont {Studenikin}},\ }\href {\doibase
  https://doi.org/10.1103/PhysRevD.100.073014} {\bibfield  {journal} {\bibinfo
  {journal} {Phys. Rev. D}\ }\textbf {\bibinfo {volume} {100}},\ \bibinfo
  {pages} {073014} (\bibinfo {year} {2019})}\BibitemShut {NoStop}%
\bibitem [{\citenamefont {Adams}(2001)}]{adamsthesis}%
  \BibitemOpen
  \bibfield  {author} {\bibinfo {author} {\bibfnamefont {J.~S.}\ \bibnamefont
  {Adams}},\ }\emph {\bibinfo {title} {Energy Deposition by Electrons in
  Superfluid Helium}},\ \href@noop {} {Ph.D. thesis},\ \bibinfo  {school}
  {Brown University} (\bibinfo {year} {2001})\BibitemShut {NoStop}%
\bibitem [{\citenamefont {Dodd}\ \emph {et~al.}(1998)\citenamefont {Dodd},
  \citenamefont {Hendry}, \citenamefont {Lawson}, \citenamefont {McClintock},\
  and\ \citenamefont {Williams}}]{Dodd:98}%
  \BibitemOpen
  \bibfield  {author} {\bibinfo {author} {\bibfnamefont {M.~E.}\ \bibnamefont
  {Dodd}}, \bibinfo {author} {\bibfnamefont {P.~C.}\ \bibnamefont {Hendry}},
  \bibinfo {author} {\bibfnamefont {N.~S.}\ \bibnamefont {Lawson}}, \bibinfo
  {author} {\bibfnamefont {P.~V.~E.}\ \bibnamefont {McClintock}}, \ and\
  \bibinfo {author} {\bibfnamefont {C.~D.~H.}\ \bibnamefont {Williams}},\
  }\href {\doibase 10.1103/PhysRevLett.81.3703} {\bibfield  {journal} {\bibinfo
   {journal} {Phys. Rev. Lett.}\ }\textbf {\bibinfo {volume} {81}},\ \bibinfo
  {pages} {3703} (\bibinfo {year} {1998})}\BibitemShut {NoStop}%
\bibitem [{\citenamefont {Forrester}\ \emph {et~al.}(2013)\citenamefont
  {Forrester}, \citenamefont {Chu},\ and\ \citenamefont
  {Williams}}]{Forrester:2013}%
  \BibitemOpen
  \bibfield  {author} {\bibinfo {author} {\bibfnamefont {A.}~\bibnamefont
  {Forrester}}, \bibinfo {author} {\bibfnamefont {H.-C.}\ \bibnamefont {Chu}},
  \ and\ \bibinfo {author} {\bibfnamefont {G.~A.}\ \bibnamefont {Williams}},\
  }\href {\doibase 10.1103/PhysRevLett.110.165303} {\bibfield  {journal}
  {\bibinfo  {journal} {Phys. Rev. Lett.}\ }\textbf {\bibinfo {volume} {110}},\
  \bibinfo {pages} {165303} (\bibinfo {year} {2013})}\BibitemShut {NoStop}%
\bibitem [{\citenamefont {Ito}\ \emph {et~al.}(2012)\citenamefont {Ito},
  \citenamefont {Clayton}, \citenamefont {Ramsey}, \citenamefont {Karcz},
  \citenamefont {Liu}, \citenamefont {Long}, \citenamefont {Reddy},\ and\
  \citenamefont {Seidel}}]{TI2012a}%
  \BibitemOpen
  \bibfield  {author} {\bibinfo {author} {\bibfnamefont {T.~M.}\ \bibnamefont
  {Ito}}, \bibinfo {author} {\bibfnamefont {S.~M.}\ \bibnamefont {Clayton}},
  \bibinfo {author} {\bibfnamefont {J.}~\bibnamefont {Ramsey}}, \bibinfo
  {author} {\bibfnamefont {M.}~\bibnamefont {Karcz}}, \bibinfo {author}
  {\bibfnamefont {C.-Y.}\ \bibnamefont {Liu}}, \bibinfo {author} {\bibfnamefont
  {J.~C.}\ \bibnamefont {Long}}, \bibinfo {author} {\bibfnamefont {T.~G.}\
  \bibnamefont {Reddy}}, \ and\ \bibinfo {author} {\bibfnamefont {G.~M.}\
  \bibnamefont {Seidel}},\ }\href {\doibase 10.1103/PhysRevA.85.042718}
  {\bibfield  {journal} {\bibinfo  {journal} {Phys. Rev. A}\ }\textbf {\bibinfo
  {volume} {85}},\ \bibinfo {pages} {042718} (\bibinfo {year}
  {2012})}\BibitemShut {NoStop}%
\bibitem [{\citenamefont {Guo}\ \emph {et~al.}(2012)\citenamefont {Guo},
  \citenamefont {Dufault}, \citenamefont {Cahn}, \citenamefont {Nikkel},
  \citenamefont {Shin},\ and\ \citenamefont {McKinsey}}]{Guo:12ji}%
  \BibitemOpen
  \bibfield  {author} {\bibinfo {author} {\bibfnamefont {W.}~\bibnamefont
  {Guo}}, \bibinfo {author} {\bibfnamefont {M.}~\bibnamefont {Dufault}},
  \bibinfo {author} {\bibfnamefont {S.~B.}\ \bibnamefont {Cahn}}, \bibinfo
  {author} {\bibfnamefont {J.~A.}\ \bibnamefont {Nikkel}}, \bibinfo {author}
  {\bibfnamefont {Y.}~\bibnamefont {Shin}}, \ and\ \bibinfo {author}
  {\bibfnamefont {D.~N.}\ \bibnamefont {McKinsey}},\ }\href
  {https://doi.org/10.1088/1748-0221/7/01/P01002} {\bibfield  {journal}
  {\bibinfo  {journal} {JINST}\ }\textbf {\bibinfo {volume} {7}},\ \bibinfo
  {pages} {P01002} (\bibinfo {year} {2012})}\BibitemShut {NoStop}%
\bibitem [{\citenamefont {Seidel}\ \emph {et~al.}(2014)\citenamefont {Seidel},
  \citenamefont {Ito}, \citenamefont {Ghosh},\ and\ \citenamefont
  {Sethumadhavan}}]{GS2014a}%
  \BibitemOpen
  \bibfield  {author} {\bibinfo {author} {\bibfnamefont {G.~M.}\ \bibnamefont
  {Seidel}}, \bibinfo {author} {\bibfnamefont {T.~M.}\ \bibnamefont {Ito}},
  \bibinfo {author} {\bibfnamefont {A.}~\bibnamefont {Ghosh}}, \ and\ \bibinfo
  {author} {\bibfnamefont {B.}~\bibnamefont {Sethumadhavan}},\ }\href {\doibase
  10.1103/PhysRevC.89.025808} {\bibfield  {journal} {\bibinfo  {journal} {Phys.
  Rev. C}\ }\textbf {\bibinfo {volume} {89}},\ \bibinfo {pages} {025808}
  (\bibinfo {year} {2014})}\BibitemShut {NoStop}%
\bibitem [{\citenamefont {Phan}\ \emph {et~al.}(2020)\citenamefont {Phan} \emph
  {et~al.}}]{NP2020a}%
  \BibitemOpen
  \bibfield  {author} {\bibinfo {author} {\bibfnamefont {N.~S.}\ \bibnamefont
  {Phan}} \emph {et~al.},\ }\href
  {https://doi-org.libproxy.berkeley.edu/10.1103/PhysRevC.102.035503}
  {\bibfield  {journal} {\bibinfo  {journal} {Phys. Rev. C}\ }\textbf {\bibinfo
  {volume} {102}},\ \bibinfo {pages} {035503} (\bibinfo {year}
  {2020})}\BibitemShut {NoStop}%
\bibitem [{\citenamefont {Seidel}(2016)}]{GS2016a}%
  \BibitemOpen
  \bibfield  {author} {\bibinfo {author} {\bibfnamefont {G.}~\bibnamefont
  {Seidel}},\ }\href
  {https://indico.physics.lbl.gov/indico/event/298/contributions/755/}
  {\enquote {\bibinfo {title} {Some properties of superfluid helium of
  relevance to dark matter detection},}\ }\bibinfo {howpublished} {{Sub-eV
  Workshop, Lawrence Berkeley National Lab}} (\bibinfo {year}
  {2016})\BibitemShut {NoStop}%
\bibitem [{\citenamefont {Seidel}(2019)}]{GS2019a}%
  \BibitemOpen
  \bibfield  {author} {\bibinfo {author} {\bibfnamefont {G.}~\bibnamefont
  {Seidel}},\ }\href
  {https://indico.ihep.ac.cn/event/10369/session/4/contribution/20} {\enquote
  {\bibinfo {title} {Properties of helium of importance for dark matter
  detection},}\ }\bibinfo {howpublished} {{Dark Matter Direct Detection
  Challenges and Opportunities, Beijing}} (\bibinfo {year} {2019})\BibitemShut
  {NoStop}%
\bibitem [{\citenamefont {Fischbach}\ \emph {et~al.}(1969)\citenamefont
  {Fischbach}, \citenamefont {Roberts},\ and\ \citenamefont
  {Hereford}}]{fischbach69}%
  \BibitemOpen
  \bibfield  {author} {\bibinfo {author} {\bibfnamefont {M.~R.}\ \bibnamefont
  {Fischbach}}, \bibinfo {author} {\bibfnamefont {H.~A.}\ \bibnamefont
  {Roberts}}, \ and\ \bibinfo {author} {\bibfnamefont {F.~L.}\ \bibnamefont
  {Hereford}},\ }\href {\doibase 10.1103/PhysRevLett.23.462} {\bibfield
  {journal} {\bibinfo  {journal} {Phys. Rev. Lett.}\ }\textbf {\bibinfo
  {volume} {23}},\ \bibinfo {pages} {462} (\bibinfo {year} {1969})}\BibitemShut
  {NoStop}%
\bibitem [{\citenamefont {McKinsey}\ \emph {et~al.}(1997)\citenamefont
  {McKinsey}, \citenamefont {Brome}, \citenamefont {Butterworth}, \citenamefont
  {Golub}, \citenamefont {Habicht}, \citenamefont {Huffman}, \citenamefont
  {Lamoreaux}, \citenamefont {Mattoni},\ and\ \citenamefont
  {Doyle}}]{MCKINSEY1997351}%
  \BibitemOpen
  \bibfield  {author} {\bibinfo {author} {\bibfnamefont {D.}~\bibnamefont
  {McKinsey}}, \bibinfo {author} {\bibfnamefont {C.~R.}\ \bibnamefont {Brome}},
  \bibinfo {author} {\bibfnamefont {J.~S.}\ \bibnamefont {Butterworth}},
  \bibinfo {author} {\bibfnamefont {R.}~\bibnamefont {Golub}}, \bibinfo
  {author} {\bibfnamefont {K.}~\bibnamefont {Habicht}}, \bibinfo {author}
  {\bibfnamefont {P.~R.}\ \bibnamefont {Huffman}}, \bibinfo {author}
  {\bibfnamefont {S.~K.}\ \bibnamefont {Lamoreaux}}, \bibinfo {author}
  {\bibfnamefont {C.~E.~H.}\ \bibnamefont {Mattoni}}, \ and\ \bibinfo {author}
  {\bibfnamefont {J.~M.}\ \bibnamefont {Doyle}},\ }\href {\doibase
  https://doi.org/10.1016/S0168-583X(97)00409-6} {\bibfield  {journal}
  {\bibinfo  {journal} {Nucl. Instrum. Methods Phys. Res. B}\ }\textbf
  {\bibinfo {volume} {132}},\ \bibinfo {pages} {351 } (\bibinfo {year}
  {1997})}\BibitemShut {NoStop}%
\bibitem [{\citenamefont {Benson}\ \emph {et~al.}(2018)\citenamefont {Benson},
  \citenamefont {Orebi~Gann},\ and\ \citenamefont {Gehman}}]{Benson2018}%
  \BibitemOpen
  \bibfield  {author} {\bibinfo {author} {\bibfnamefont {C.}~\bibnamefont
  {Benson}}, \bibinfo {author} {\bibfnamefont {G.~D.}\ \bibnamefont
  {Orebi~Gann}}, \ and\ \bibinfo {author} {\bibfnamefont {V.}~\bibnamefont
  {Gehman}},\ }\href {\doibase 10.1140/epjc/s10052-018-5807-z} {\bibfield
  {journal} {\bibinfo  {journal} {Eur. Phys. J. C}\ }\textbf {\bibinfo {volume}
  {78}},\ \bibinfo {pages} {329} (\bibinfo {year} {2018})}\BibitemShut
  {NoStop}%
\bibitem [{\citenamefont {Zhang}\ \emph {et~al.}(2016)\citenamefont {Zhang},
  \citenamefont {Lin}, \citenamefont {Mikhailik},\ and\ \citenamefont
  {Kraus}}]{ZHANG2016}%
  \BibitemOpen
  \bibfield  {author} {\bibinfo {author} {\bibfnamefont {X.}~\bibnamefont
  {Zhang}}, \bibinfo {author} {\bibfnamefont {J.}~\bibnamefont {Lin}}, \bibinfo
  {author} {\bibfnamefont {V.}~\bibnamefont {Mikhailik}}, \ and\ \bibinfo
  {author} {\bibfnamefont {H.}~\bibnamefont {Kraus}},\ }\href {\doibase
  https://doi.org/10.1016/j.astropartphys.2016.02.007} {\bibfield  {journal}
  {\bibinfo  {journal} {Astropart. Phys.}\ }\textbf {\bibinfo {volume} {79}},\
  \bibinfo {pages} {31 } (\bibinfo {year} {2016})}\BibitemShut {NoStop}%
\bibitem [{\citenamefont {Hyodo}\ and\ \citenamefont
  {Nagai}(1980)}]{Hyodo1980}%
  \BibitemOpen
  \bibfield  {author} {\bibinfo {author} {\bibfnamefont {S.}~\bibnamefont
  {Hyodo}}\ and\ \bibinfo {author} {\bibfnamefont {A.}~\bibnamefont {Nagai}},\
  }\href@noop {} {\bibfield  {journal} {\bibinfo  {journal} {Journal of the
  Faculty of Engineering, University of Tokyo, Series A}\ }\textbf {\bibinfo
  {volume} {18}} (\bibinfo {year} {1980})}\BibitemShut {NoStop}%
\bibitem [{\citenamefont {Agostinelli}\ \emph {et~al.}(2003)\citenamefont
  {Agostinelli} \emph {et~al.}}]{Agostinelli:03nimpra}%
  \BibitemOpen
  \bibfield  {author} {\bibinfo {author} {\bibfnamefont {S.}~\bibnamefont
  {Agostinelli}} \emph {et~al.},\ }\href {\doibase
  10.1016/s0168-9002(03)01368-8} {\bibfield  {journal} {\bibinfo  {journal}
  {Nucl. Instr. Meth. Phys. Res. A}\ }\textbf {\bibinfo {volume} {506}},\
  \bibinfo {pages} {250} (\bibinfo {year} {2003})}\BibitemShut {NoStop}%
\bibitem [{\citenamefont {Allison}\ \emph {et~al.}(2006)\citenamefont {Allison}
  \emph {et~al.}}]{Allison:IEEE}%
  \BibitemOpen
  \bibfield  {author} {\bibinfo {author} {\bibfnamefont {J.}~\bibnamefont
  {Allison}} \emph {et~al.},\ }\href
  {http://dx.doi.org/10.1109/TNS.2006.869826} {\bibfield  {journal} {\bibinfo
  {journal} {IEEE Transactions on Nuclear Science}\ }\textbf {\bibinfo {volume}
  {53}},\ \bibinfo {pages} {270} (\bibinfo {year} {2006})}\BibitemShut
  {NoStop}%
\bibitem [{\citenamefont {Allison}\ \emph {et~al.}(2016)\citenamefont {Allison}
  \emph {et~al.}}]{Allison:16nimpra}%
  \BibitemOpen
  \bibfield  {author} {\bibinfo {author} {\bibfnamefont {J.}~\bibnamefont
  {Allison}} \emph {et~al.},\ }\href
  {http://dx.doi.org/10.1016/j.nima.2016.06.125} {\bibfield  {journal}
  {\bibinfo  {journal} {Nucl. Instr. Meth. Phys. Res. A}\ }\textbf {\bibinfo
  {volume} {835}},\ \bibinfo {pages} {186} (\bibinfo {year}
  {2016})}\BibitemShut {NoStop}%
\bibitem [{\citenamefont {{Geant4 Collaboration}}(2019)}]{G4_physics}%
  \BibitemOpen
  \bibfield  {author} {\bibinfo {author} {\bibnamefont {{Geant4
  Collaboration}}},\ }\href@noop {} {\enquote {\bibinfo {title} {{Guide For
  Physics Lists Release 10.5}},}\ }\bibinfo {howpublished}
  {\url{https://geant4-userdoc.web.cern.ch/UsersGuides/PhysicsListGuide/BackupVersions/V10.5-2.0/html/index.html}}
  (\bibinfo {year} {2019})\BibitemShut {NoStop}%
\bibitem [{\citenamefont {Liskien}\ and\ \citenamefont
  {Paulsen}(1973)}]{LISKIEN1973569}%
  \BibitemOpen
  \bibfield  {author} {\bibinfo {author} {\bibfnamefont {H.}~\bibnamefont
  {Liskien}}\ and\ \bibinfo {author} {\bibfnamefont {A.}~\bibnamefont
  {Paulsen}},\ }\href {\doibase https://doi.org/10.1016/S0092-640X(73)80081-6}
  {\bibfield  {journal} {\bibinfo  {journal} {At. Data Nucl. Data Tables}\
  }\textbf {\bibinfo {volume} {11}},\ \bibinfo {pages} {569 } (\bibinfo {year}
  {1973})}\BibitemShut {NoStop}%
\bibitem [{\citenamefont {Berger}\ \emph {et~al.}(2017)\citenamefont {Berger},
  \citenamefont {Coursey}, \citenamefont {Zucker},\ and\ \citenamefont
  {Chang}}]{NIST_STAR}%
  \BibitemOpen
  \bibfield  {author} {\bibinfo {author} {\bibfnamefont {M.~J.}\ \bibnamefont
  {Berger}}, \bibinfo {author} {\bibfnamefont {J.~S.}\ \bibnamefont {Coursey}},
  \bibinfo {author} {\bibfnamefont {M.~A.}\ \bibnamefont {Zucker}}, \ and\
  \bibinfo {author} {\bibfnamefont {J.}~\bibnamefont {Chang}},\ }\href
  {http://physics.nist.gov/Star} {\enquote {\bibinfo {title} {{ESTAR, PSTAR,
  and ASTAR}: Computer programs for calculating stopping-power and range tables
  for electrons, protons, and helium ions},}\ } (\bibinfo {year}
  {2017})\BibitemShut {NoStop}%
\bibitem [{\citenamefont {King}\ and\ \citenamefont {Voltz}(1966)}]{King:66}%
  \BibitemOpen
  \bibfield  {author} {\bibinfo {author} {\bibfnamefont {T.~A.}\ \bibnamefont
  {King}}\ and\ \bibinfo {author} {\bibfnamefont {R.}~\bibnamefont {Voltz}},\
  }\href@noop {} {\bibfield  {journal} {\bibinfo  {journal} {Proc. R. Soc.
  Lond.}\ }\textbf {\bibinfo {volume} {289}},\ \bibinfo {pages} {424} (\bibinfo
  {year} {1966})}\BibitemShut {NoStop}%
\bibitem [{\citenamefont {Rudd}\ \emph {et~al.}(1992)\citenamefont {Rudd},
  \citenamefont {Kim}, \citenamefont {Madison},\ and\ \citenamefont
  {Gay}}]{MR1992a}%
  \BibitemOpen
  \bibfield  {author} {\bibinfo {author} {\bibfnamefont {M.~E.}\ \bibnamefont
  {Rudd}}, \bibinfo {author} {\bibfnamefont {Y.-K.}\ \bibnamefont {Kim}},
  \bibinfo {author} {\bibfnamefont {D.~H.}\ \bibnamefont {Madison}}, \ and\
  \bibinfo {author} {\bibfnamefont {T.~J.}\ \bibnamefont {Gay}},\ }\href@noop
  {} {\bibfield  {journal} {\bibinfo  {journal} {Rev. Mod. Phys.}\ }\textbf
  {\bibinfo {volume} {64}},\ \bibinfo {pages} {441} (\bibinfo {year}
  {1992})}\BibitemShut {NoStop}%
\bibitem [{\citenamefont {Keto}\ \emph {et~al.}(1974)\citenamefont {Keto},
  \citenamefont {Soley}, \citenamefont {Stockton},\ and\ \citenamefont
  {Fitzsimmons}}]{Keto:1974}%
  \BibitemOpen
  \bibfield  {author} {\bibinfo {author} {\bibfnamefont {J.~W.}\ \bibnamefont
  {Keto}}, \bibinfo {author} {\bibfnamefont {F.~J.}\ \bibnamefont {Soley}},
  \bibinfo {author} {\bibfnamefont {M.}~\bibnamefont {Stockton}}, \ and\
  \bibinfo {author} {\bibfnamefont {W.~A.}\ \bibnamefont {Fitzsimmons}},\
  }\href {\doibase 10.1103/PhysRevA.10.887} {\bibfield  {journal} {\bibinfo
  {journal} {Phys. Rev. A}\ }\textbf {\bibinfo {volume} {10}},\ \bibinfo
  {pages} {887} (\bibinfo {year} {1974})}\BibitemShut {NoStop}%
\bibitem [{\citenamefont {Roberts}\ and\ \citenamefont
  {Hereford}(1973)}]{Roberts:1973}%
  \BibitemOpen
  \bibfield  {author} {\bibinfo {author} {\bibfnamefont {H.~A.}\ \bibnamefont
  {Roberts}}\ and\ \bibinfo {author} {\bibfnamefont {F.~L.}\ \bibnamefont
  {Hereford}},\ }\href {\doibase 10.1103/PhysRevA.7.284} {\bibfield  {journal}
  {\bibinfo  {journal} {Phys. Rev. A}\ }\textbf {\bibinfo {volume} {7}},\
  \bibinfo {pages} {284} (\bibinfo {year} {1973})}\BibitemShut {NoStop}%
\bibitem [{\citenamefont {Segreto}(2015)}]{Segreto_2015}%
  \BibitemOpen
  \bibfield  {author} {\bibinfo {author} {\bibfnamefont {E.}~\bibnamefont
  {Segreto}},\ }\href {\doibase 10.1103/physrevc.91.035503} {\bibfield
  {journal} {\bibinfo  {journal} {Phys. Rev. C}\ }\textbf {\bibinfo {volume}
  {91}},\ \bibinfo {pages} {035503} (\bibinfo {year} {2015})}\BibitemShut
  {NoStop}%
\end{thebibliography}%

\end{document}